\documentclass[sn-mathphys]{sn-jnl-ar}

\jyear{2021}%

\theoremstyle{thmstyleone}%
\theoremstyle{thmstyletwo}%

\theoremstyle{thmstylethree}%

\begin{document}

\title[PRIDE JUICE]{Planetary Radio Interferometry and Doppler Experiment (PRIDE) of the JUICE mission}


\author*[1,2]{\fnm{Leonid} \sur{I. Gurvits}}\email{lgurvits@jive.eu, ORCID: 0000-0002-0694-2459}

\author[1]{\fnm{Giuseppe} \sur{Cim\`{o}}}
\email{(cimo@jive.eu, ORCID: 0000-0002-1167-7565)}
\equalcont{These authors contributed equally to this work.}

\author[2]{\fnm{Dominic} \sur{Dirkx}}\email{(d.dirkx@tudelft.nl, ORCID: 0000-0003-2069-0603)}
\equalcont{These authors contributed equally to this work.}

\author[2]{\fnm{Vidhya} \sur{Pallichadath}}\email{(v.pallichadath@tudelft.nl, ORCID: 0000-0003-0678-4814)}
\equalcont{These authors contributed equally to this work.}

\author[4]{\fnm{Alexander} \sur{Akins}}\email{(alexander.akins@jpl.nasa.gov, ORCID: 0000-0001-8379-1909)}
\equalcont{These authors contributed equally to this work.}

\author[3]{\fnm{Nicolas} \sur{Altobelli}}\email{(nicolas.altobelli@esa.int, ORCID: 0000-0003-4244-6302)}
\equalcont{These authors contributed equally to this work.}

\author[4]{\fnm{Tatiana~M.} \sur{Bocanegra-Bahamon}}\email{(tatiana.m.bocanegra.bahamon@jpl.nasa.gov, ORCID: 0000-0001-8338-8650)}
\equalcont{These authors contributed equally to this work.}

\author[2]{\fnm{St\'{e}phanie} \sur{M.~Cazaux}}\email{(s.m.cazaux@tudelft.nl, ORCID: 0000-0003-0320-3578)}
\equalcont{These authors contributed equally to this work.}

\author[5]{\fnm{Patrick} \sur{Charlot}}\email{(patrick.charlot@u-bordeaux.fr, ORCID: 0000-0002-9142-716X)}
\equalcont{These authors contributed equally to this work.}

\author[6]{\fnm{Dmitry~A.} \sur{Duev}}\email{(duev@wandb.ai, ORCID: 0000-0001-5060-8733)}
\equalcont{These authors contributed equally to this work.}

\author[2]{\fnm{Marie~S.} \sur{Fayolle}}\email{(m.s.fayolle-chambe@tudelft.nl, ORCID: 0000-0003-4407-5031)}
\equalcont{These authors contributed equally to this work.}

\author[7]{\fnm{Judit} \sur{Fogasy}}\email{(fogasy.judit@csfk.org, ORCID: 0000-0003-0003-000X)}
\equalcont{These authors contributed equally to this work.}

\author[7]{\fnm{S\'{a}ndor} \sur{Frey}}\email{(frey.sandor@csfk.org, ORCID: 0000-0003-3079-1889}
\equalcont{These authors contributed equally to this work.}

\author[8]{\fnm{Valery} \sur{Lainey}}\email{(lainey@imcce.fr, ORCID: 0000-0003-1618-4281)}
\equalcont{These authors contributed equally to this work.}

\author[9]{\fnm{Guifr\'{e}} \sur{Molera~Calv\'{e}s}}\email{(guifre.moleracalves@utas.edu.au, ORCID: 0000-0001-8819-0651)}
\equalcont{These authors contributed equally to this work.}

\author[7]{\fnm{Krisztina} \sur{Perger}}\email{(perger.krisztina@csfk.org, ORCID: 0000-0002-6044-6069)}
\equalcont{These authors contributed equally to this work.}

\author[1]{\fnm{Sergey~V.} \sur{Pogrebenko}}\email{spogrebenko@gmail.com}
\equalcont{These authors contributed equally to this work.}

\author[1]{\fnm{N.~Masdiana} \sur{Md~Said}}\email{(said@jive.eu, ORCID: 0000-0002-7147-7039)}
\equalcont{These authors contributed equally to this work.}

\author[3]{\fnm{Claire} \sur{Vallat}}\email{claire.vallat@esa.int}
\equalcont{These authors contributed equally to this work.}

\author[2]{\fnm{Bert~L.A.} \sur{Vermeersen}}\email{(l.l.a.vermeersen@tudelft.nl, ORCID: 0000-0002-6329-5972)}
\equalcont{These authors contributed equally to this work.}

\author[2]{\fnm{Pieter~N.A.M.} \sur{Visser}}\email{(p.n.a.m.visser@tudelft.nl, ORCID: 0000-0002-2018-7373)}
\equalcont{These authors contributed equally to this work.}

\author[4]{\fnm{Kuo-Nung} \sur{Wang}}
\email{(kuo-nung.wang@jpl.nasa.gov, ORCID: 0000-0002-9599-6132)}
\equalcont{These authors contributed equally to this work.}

\author[10]{\fnm{Konrad} \sur{Willner}}\email{(konrad.willner@dlr.de, ORCID: 0000-0002-5437-8477)}
\equalcont{These authors contributed equally to this work.}



\affil*[1]{\orgname{Joint Institute for VLBI ERIC}, \orgaddress{\street{Oude Hoogeveensedijk~4}, \city{Dwingeloo}, \postcode{7991~PD}, \country{The Netherlands}}}

\affil[2]{\orgdiv{Faculty of Aerospace Engineering}, \orgname{Delft University of Technology}, \orgaddress{\street{Kluyverweg 1}, \city{Delft}, \postcode{2629 HS}, \country{The Netherlands}}}

\affil[3]{\orgdiv{European Space Astronomy Centre}, \orgname{European Space Agency}, \orgaddress{\street{Camino Bajo del Castillo s/n Villafranca del Castillo}, \city{Villanueva de la Ca\~{n}ada (Madrid)}, \postcode{28692}, \country{Spain}}}

\affil[4]{\orgdiv{Jet Propulsion Laboratory}, \orgname{California Institute of Technology}, \orgaddress{\street{4800 Oak Grove Dr.}, \city{Pasadena}, \postcode{91109}, \state{CA}, \country{USA}}}

\affil[5]{\orgdiv{Laboratoire d’astrophysique de Bordeaux}, \orgname{Univ. Bordeaux, CNRS}, \orgaddress{\street{B18N, All\'{e}e Geoffroy Saint-Hilaire}, \postcode{33615} \city{Pessac}, \country{France}}}

\affil[6]{\orgname{Weights \& Biases Inc.}, \orgaddress{\street{1479 Folsom St.}, \city{San Francisco}, \postcode{94103}, \state{CA}, \country{USA}}}

\affil[7]{\orgdiv{Konkoly Observatory}, \orgname{HUN-REN Research Centre for Astronomy and Earth Sciences, MTA Centre of Excellence}, \orgaddress{\street{Konkoly Thege M. \'{u}t 15-17}, \city{Budapest}, \postcode{H--1121}, \country{Hungary}}}

\affil[8]{\orgdiv{IMCCE}, \orgname{Observatoire de Paris}, \orgaddress{\street{Street}, \city{Paris}, \postcode{75014}, \country{France}}}

\affil[9]{\orgdiv{Physics discipline, School of Natural Sciences}, \orgname{University of Tasmania}, \orgaddress{\street{Street}, \city{Hobart}, \postcode{TAS~7000}, \state{Tasmania}, \country{Australia}}}

\affil[10]{\orgdiv{Institute of Planetary Research, Planetary Geodesy}, \orgname{German Aerospace Center (DLR)}, \orgaddress{\street{Rutherfordstr. 2}, \city{Berlin}, \postcode{12489}, \country{Germany}}}



\abstract{Planetary Radio Interferometry and Doppler Experiment (PRIDE) is a multi-purpose experimental technique aimed at enhancing the science return of planetary missions. The technique exploits the science payload and spacecraft service systems without requiring a dedicated onboard instrumentation or imposing on the existing instrumentation any special for PRIDE requirements. PRIDE is based on the near-field phase-referencing Very Long Baseline Interferometry (VLBI) and evaluation of the Doppler shift of the radio signal transmitted by spacecraft by observing it with multiple Earth-based radio telescopes. The methodology of PRIDE has been developed initially at the Joint Institute for VLBI ERIC (JIVE) for tracking the ESA’s Huygens Probe during its descent in the atmosphere of Titan in 2005. From that point on, the technique has been demonstrated for various planetary and other space science missions. The estimates of lateral position of the target spacecraft are done using the phase-referencing VLBI technique. Together with radial Doppler estimates, these observables can be used for a variety of applications, including improving the knowledge of the spacecraft state vector. The PRIDE measurements can be applied to a broad scope of research fields including studies of atmospheres through the use of radio occultations, the improvement of planetary and satellite ephemerides, as well as gravity field parameters and other geodetic properties of interest, and estimations of interplanetary plasma properties. This paper presents the implementation of PRIDE as a component of the ESA's Jupiter Icy Moons Explorer (JUICE) mission.
}




\maketitle

\section{Introduction}
\label{s:intro}




A radio astronomy technique of Very Long Baseline Interferometry (VLBI) has been first demonstrated independently by three groups in the US and one in Canada in 1967 \cite[see][section~1.3.14 and references therein]{TMS-2017}. This technique is distinguished by its high angular resolution denoted here as $\theta$, which is defined by a simple expression $\theta \simeq \lambda/B$, where $\lambda$ is the observing wavelength, and $B$ is the size of the telescope's aperture. The latter is the diameter of the antenna for what is called in the radio astronomy slang ``single dish'' telescopes, or, for interferometers, the length of baseline vector projected on the plane perpendicular to the direction of the celestial source (the so-called image plane). By the middle of the 1960s, the major driver of this invention has been the accumulated understanding that many astrophysical processes in galactic and extragalactic objects are confined within compact areas, and their investigation requires angular resolutions beyond the reach of single dish antennas or even conventional interferometers with physically connected elements and baselines up to several kilometers \cite[e.g.,][section~1.3]{TMS-2017}. In the context of this special issue of \textit{Space Science Reviews} dedicated to the JUICE mission, it is interesting to note that, in January 1967, one of the targets in the pioneering VLBI observations at 18~MHz ($\lambda = 16.6$~m) was Jupiter \cite{Brown+1968ApL}, the strongest celestial radio source at decameter wavelengths. 

The record-holding angular resolution of VLBI brings about not only the ability to resolve and reconstruct the distribution of brightness in the observing target source. The technique also provides most accurate measurements of the celestial position of a radio emitting source. This astrometric application of VLBI is described in \cite[][Chapter~12]{TMS-2017} and, to the extent of relevance to the current paper, in section~\ref{s:ph-ref-catalog}. It did not take long to consider applications of the astrometric VLBI to determinations of celestial position of spacecraft. Indeed, a radio-emitting spacecraft (more specifically -- its transmitting antenna) is an ideal point-like source for VLBI observations. This idea has been considered in the beginning of the 1970s. One of the early demonstrations of VLBI tracking of spaceborne transmitters was conducted in 1973–1974 with Apollo Lunar Surface Experiments Package (ALSEP) instruments placed on the surface of the Moon by NASA's Apollo--12,14,15,16,17 missions  \cite{King+1976Alsep}. The observations were conducted at the frequency of 2.3~GHz in the so called differential interferometry mode \cite{Counselman+1972Sci}. In combination with laser ranging as a part of ALSEP, these observations yielded most accurate estimates of parameters in models of the Lunar orbit and libration and selenodetic coordinates of the radio transmitters and retroreflectors.

First practical demonstrations of tracking of spaceborne transmitters, principally based on the VLBI methodology, deviated from astronomical VLBI in several important details, most notably in using the differential delay between a background celestial radio source (usually a  quasar) and a spacecraft as the main measurable \cite{Ondrasik+1971,Melbourne+1977,Border+1982}. The technique has been successfully demonstrated for NASA's Voyager spacecraft under the name ``delta VLBI'' \cite{Brunn+1978}. Later the technique was renamed $\Delta$DOR (Delta-Differential One-way Ranging, \cite[][and references therein]{Curkendall+2013}) and has become a mainstay technique for precise state vector determination of deep space missions. While resembling astronomical VLBI instrumentation in principle, $\Delta$DOR requires a spacecraft transponder to modulate a carrier signal with a series of pure tones of escalating frequency relative to the carrier. A nominal configuration of $\Delta$DOR requires only two Earth-based stations (telescopes). The technique provides the accuracies of determination of the celestial position of spacecraft approaching one nanoradian ($\sim 200$~$\mu$as). 

Parallel developments in astronomical VLBI have resulted in the emergence of a method of phase-referencing \cite[][section~12.2.3; also this paper~\ref{ss:nfVLBI} and \ref{s:ph-ref-catalog}]{TMS-2017} enabling measurements of relative positions of closely spaced sources with the accuracy that corresponds to the angular resolution of the interferometer. In applications to spacecraft tracking, this technique principally is not inferior to $\Delta$DOR in terms of angular positioning accuracy (in some special configurations it can be superior), and is less restrictive regarding the type of the spacecraft radio signal. It also might offer some operational benefits as the number of VLBI-capable telescopes in the world is significantly larger than dedicated deep space tracking stations. Further discussions on various flavours of interferometric determinations of angular position of spacecraft are given in \cite{Lanyi+2007}.

Various VLBI observations of spacecraft were conducted in the second half of the 1970s, during the first decade of the VLBI observations in the interests of ``traditional'' astronomy. One of them was the NASA's Pioneer Venus VLBI experiment \cite{Pioneer-V-1979}. It was followed by another Venusian mission, VEGA, in which VLBI measurements played the key role in measuring the wind parameters in the planetary atmosphere via tracking two balloons \cite{VEGA-1986Venus, VEGA-RZS-1990}. Two VEGA spacecraft moved on from Venus to the encounter with the comet Halley, and their VLBI tracking, as a part of the Pathfinder experiment, conducted jointly by the European, Soviet and US space agencies, helped to navigate the ESA's Giotto spacecraft with the precision of $\sim 50$~km toward a fly-by of the comet's nucleus \cite[][and references therein]{Pathfinder-1986}. 

The Chinese deep space exploration program adopted VLBI tracking as one of the main means of spacecraft state vector determination for the Lunar projects Chang'E \cite[][and references thererin]{Changhe-R-2010S,Change1-2-2012S}, including a successful demonstration of the so called in-beam VLBI phase-referencing for the Chang'E-3 mission \cite{ChangE3-2015}. For these Lunar missions, VLBI tracking was used for orbiters, landers and rovers. Further development of this technology enabled efficient VLBI tracking of the first Chinese Martian mission Tianwen–1 in 2020--2021 \cite{Tianwen-2022SSPMA}. Another example of Lunar mission employing VLBI tracking was the Japanese Lunar gravimetry mission SELENE (Kaguya) with its dedicated small VLBI ``beacon'' satellite Ouna \cite{SELENE-VERA-2011}. The applicability of VLBI tracking has also been demonstrated for the Japanese small scale solar sail demonstrator IKAROS that has completed its nominal mission with a Venus fly-by in 2010 \cite{IKAROS-2011}. The 10-telescope Very Long Baseline Array (VLBA) in the US has demonstrated its might in VLBI tracking of the NASA's Mars missions MRO (Mars Reconnaissance Orbiter) and Odyssey \cite{VLBA-Mars-2015}, as well as the Cassini and Juno spacecraft \cite{Jones+2011Cassini, Park+2021AJ}. 

The experiment described in this publication, Planetary Radio Interferometry and Doppler Experiment (PRIDE) is a direct descendant of the ad hoc VLBI tracking of the ESA's Huygens Titan Probe in 2005 \cite{SVP+2004}. Though the tracking was not a part of the nominal Cassini--Huygens mission, it had been initiated in 2003, six years after the mission launch. In spite of the late addition to the Huygens mission science suite, this tracking exercise has assisted in achieving the mission's science goals \cite{Lebreton+2005,DWE-2005Natur,Witasse+2006JGRE}. Further development of the methodology at the Joint Institute for VLBI ERIC (JIVE) and Delft University of Technology (TU Delft) enabled multi-purpose tracking of the ESA's Mars Express (MEX) \cite{Duev+2016,Bocanegra+2018} and Venus Express (VEX) \cite{Bocanegra+2019} missions. The technique also supported high-precision orbit determination of the Space VLBI mission RadioAstron on its high-eccentricity 9-day geocentric orbit with the apogee of $\sim 350,000$~km \cite{Duev+2015}.

\begin{figure}[!hbt]
    \centering
     \includegraphics[width=\columnwidth]{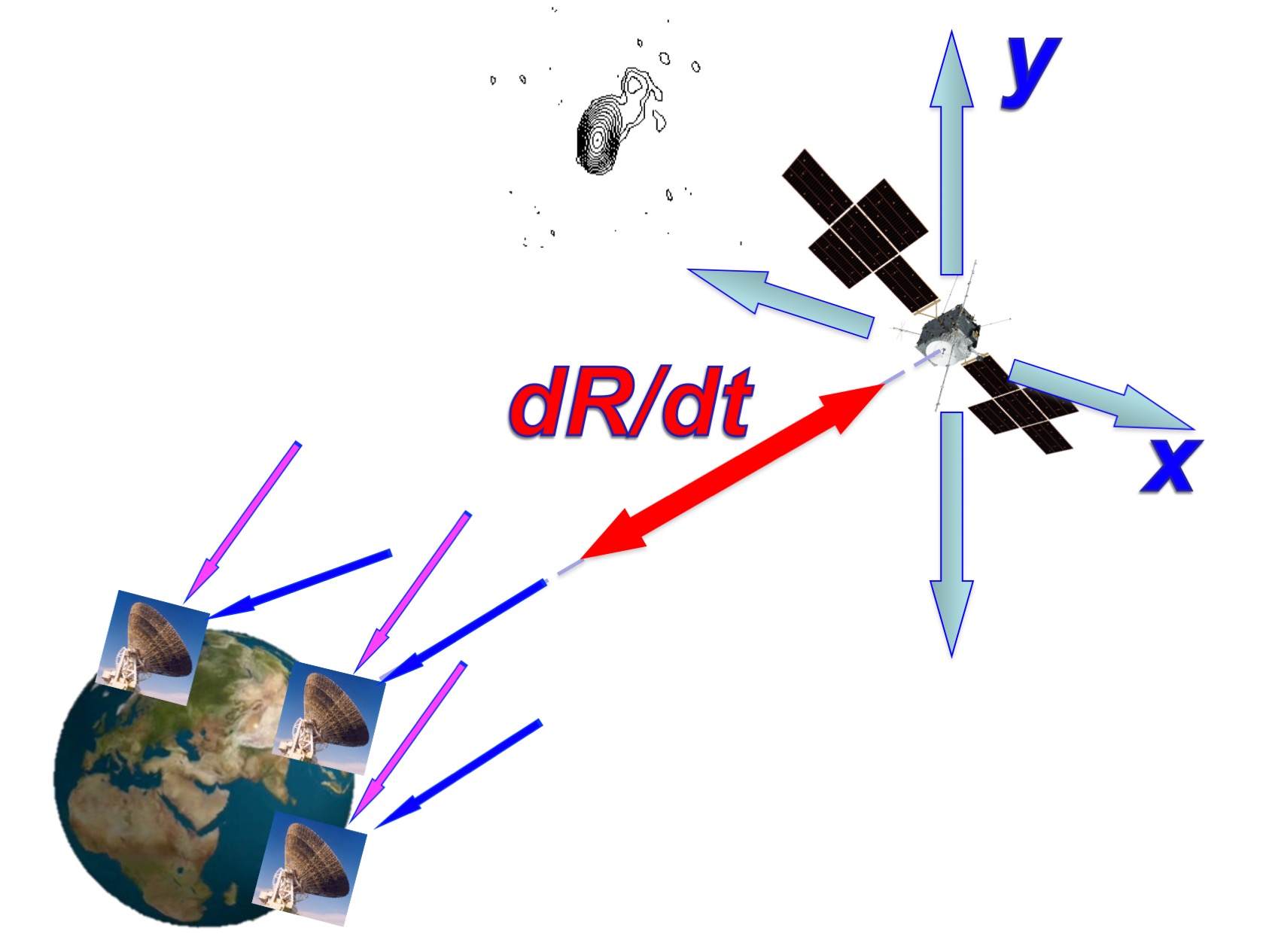}
    \caption{A principle configuration of PRIDE. Ground-based telescopes (three in the shown example) observe intermittently the spacecraft and a distant background compact natural source (shown at the top of the figure) which serves as a reference source. Ray paths from the reference source shown as magenta arrows are considered to be parallel as the source is in the far field of the interferometer. The ray paths from the spacecraft shown as blue arrows are not parallel as the spacecraft is in the near field. PRIDE observables are the lateral celestial coordinates $x$ and $y$, obtained with VLBI, and radial (Doppler) velocity $dR/dt$.}
    \label{fig:PRIDE-gen}
\end{figure}

The essence of PRIDE is a modification of the VLBI technique in which a typical for ``traditional'' astronomical interferometry assumption on the infinitely large distance to the source does not hold. This deviation effectively means that the wave front of the source's emission arriving to the interferometer's elements is not planar, i.e. directions from the interferometer's elements to the source cannot be considered parallel, Fig.~\ref{fig:PRIDE-gen}. The latter is the essential feature of the so called near field (NF) VLBI. The border between the near and far field is defined by the distance $R_{\rm nf}$ to the source at which the divergence of the wavefront from a plane is comparable to or larger than the wavelength $\lambda$. In ``traditional'' far field VLBI, it is conventionally accepted that $R_{\rm nf} = B^{2}/\lambda$ \cite[][section 15.1.3]{TMS-2017}. It is justifiable to call this latter value ``Fraunhofer criterium'' as it is a half of the so called Fraunhofer distance, $2B^{2}/\lambda$ \cite{Krish-Ram-2017}. Table~\ref{tab:nfVLBI} illustrates the practical radii of NF VLBI for two JUICE communication bands and various configurations of ground-based interferometric facilities. For typical VLBI facility with the baseline of the order of 1000 km (e.g., the Europe-located part of EVN (European VLBI Network); see Table~\ref{tab:vlbi-arrays} and corresponding footnotes), the entire Solar System is within the NF VLBI radius.

\begin{table}[h]
\caption{NF VLBI radii for various ground-based model interferometric (VLBI) facilities at two JUICE communications frequency bands, $X$- and $Ka$-bands.}
\begin{tabular}{@{}lllll@{}}
\toprule
Baseline length & 10 km  & 100 km & 1000 km & $10^{4}$ km \\
Model VLBI facility &         & SKA$^{\rm a}$ core & EVN$_{\rm Europe}$ & EVN$_{\rm Global}$ \\
\midrule
$\lambda=3.6$~cm, $X$-band    & $3\times10^{6}$ km & 2 AU  & 200 AU  & 0.1 pc \\
$\lambda=9$~mm, $Ka$-band    & $12\times10^{6}$ km  & 8 AU & 800 AU  & 0.4 pc\\
\botrule
\end{tabular}
\footnotetext{$^{\rm a}$Square Kilometre Array.}
\label{tab:nfVLBI}
\end{table}

PRIDE operates with the radio signal generated by the spacecraft transmitters. Such a signal contains a narrow sinusoidal carrier. Measuring the deviation of this carrier frequency from the nominal value enables PRIDE to estimate the radial velocity of the spacecraft via measurements of the carrier frequency Doppler shift.

The main task of PRIDE is an enhancement of other JUICE experiments by providing measurements of the lateral celestial position and radial velocity of the spacecraft which help to improve the accuracy of the spacecraft state vector determination for various applications. One of them is an improvement of the ephemerides of Jupiter and Jovian system bodies. Such the use of VLBI has been demonstrated for the system of Saturn with the Cassini VLBI observations \cite{Jones+2011Cassini} and presented by the PRIDE team for upcoming Jovian missions in \cite{Dirkx+2016, Dirkx+2017, Fayolle+2022PSS}, most notably the JUICE mission. In addition, JIVE will contribute to the JUICE mission science by observing JUICE spacecraft radio occultations (section~\ref{s:occult}). All these scientific applications of PRIDE are synergistic to the overall JUICE mission tasks described in this Special Issue of Space Science Reviews. In particular, PRIDE will provide supplementary measurements of the spacecraft differential lateral position relative to the ICRF (International Celestial Reference Frame) background extragalactic radio sources with the accuracy of 100--10~$\mu$as (1$\sigma$ RMS) over integration time 60--1000~s. These measurements will contribute to characterisation of the the interiors \cite{VanHoolst+2023SSR} as well as surfaces and near-surface atmospheres \cite{Tosi+2023SSR} of Ganymede, Europa and Callisto. PRIDE evaluation of the radio wave propagation medium described in section~\ref{s:plasma} will contribute to the JUICE studies of magnetosphere and interplanetary plasma \cite{Masters+2023SSR}.

Two experiments of the JUICE mission will address the mission science themes by means of radio science methods, the Gravity and Geophysics of Jupiter and Galilean Moons (3GM, \cite{Iess+2023SSR}) experiment and PRIDE. These two experiments overlap partially in covering several topics of the JUICE science program as well as in some technical implementation issues. In order to optimise use of the mission resources and maximise the mission science return, special attention was given to coordination between 3GM and PRIDE. As a part of this coordination, it was agreed that 3GM will take the prime responsibility for investigation of geophysical properties of Ganymede, including its gravity field and interior. It will also lead radio occultation sounding of the atmospheres of Jupiter and its moons. PRIDE will participate in all these investigations by providing additional measurements using Earth-based VLBI radio telescopes. At the same time, the prime focus of PRIDE will be the improvement of ephemerides of the Galilean moons. It will also lead the coordinated efforts in using bistatic or ``multi-static'' measurements for characterisation of the moons surfaces, including observations around radio occultations. Several other scientific topics will be addressed by 3GM and JIVE jointly. In any case, all these and other radio science measurements will be synergistic with investigations by other JUICE instruments in achieving the mission scientific objectives.

This paper is organised as following. Section~\ref{s:PRIDE-glance} describes the instrumentation and infrastructure of PRIDE and its interface to the JUICE mission instrumental and operational components. Section~\ref{s:PRIDE-proc} presents the algorithms and their software implementation of the PRIDE VLBI and Doppler data processing. In section~\ref{s:ph-ref-catalog} we discuss the requirements to and composition of catalogues of natural celestial background reference radio sources, including those which form the International Celestial Reference Frame (ICRF). Section~\ref{s:schd} is devoted to the observations scheduling issues; this topic is the central operational element of the interaction between the JUICE mission operations and operations of the globally distributed networks of VLBI radio telescopes. In section~\ref{s:state-v} we discuss the use of PRIDE measurements for the spacecraft state vector determination. Section~\ref{s:ephemer} describes the contribution of PRIDE into improvement of Jovian moon's ephemerides. Sections~\ref{s:occult} and \ref{s:plasma} present ad hoc applications of PRIDE for radio occultation experiments with JUICE and diagnostics of the interplanetary plasma, respectively. In section~\ref{s:Eu-Clip} potential synergy between JUICE and Europa Clipper observations is discussed. Finally, section~\ref{s:conclu} summarises the outlook of PRIDE contribution to the JUICE mission science.  

\section{PRIDE instrumentation and infrastructure}
\label{s:PRIDE-glance}

PRIDE is conceived as an ad hoc enhancement of the mission science output relying on the available onboard and Earth-based instrumentation and infrastructures. Both these major components of PRIDE have been designed and built for other than PRIDE purposes. During the mission in-flight operations, these components operate in order to implement their prime purposes. While for the JUICE onboard instrumentation this purpose is obviously defined by the mission goals, the Earth-based global network of VLBI radio telescopes pursue its own science agenda. Therefore, for PRIDE, it is essential to define and exploit the interface between these two very diverse components in the most efficient way.

\subsection{PRIDE interface to the mission on-board instrumentation and Earth-based segment}
\label{ss:PRIDE-onboard}


The JUICE mission is set to explore Jupiter and its icy moons, Ganymede, Europa, and Callisto, with the emphasis on Ganymede from the polar orbit around this moon in the final phase of the mission starting in 2034 \citep{dougherty2011juice, grasset2013jupiter, Witasse+2023SSR}. The cruise phase, which will last for nearly 8 years before the {Jupiter Orbit Insertion} (JOI), will include flybys of the Moon and Earth in August 2024, two flybys of Earth in September 2026 and January 2029, and a flyby of Venus in August 2025 (Consolidated Report on Mission Analysis (CReMA) 5.0b23 \citep{boutonnet2017juice, Boutonnet+2023SSR}). 
 
 PRIDE will augment the determination of the spacecraft state vector by the mission's nominal means and other science instruments by adding VLBI (the lateral celestial position) and radial velocity (Doppler) measurements using the JUICE radio communication system. A redundant set of transponders that employ X-band for the uplink as well as X- and Ka-band for the downlink compose the communications subsystem of the spacecraft. A Ka/Ka link (KAT) at 32--34~GHz, a part of the JUICE radio science experiment 3GM will be used for Doppler (range rate) and range measurements in the two-way (closed loop) mode \cite{Iess+2023SSR}. Another part of the 3GM instrument, the Ultra Stable Oscillator (USO) will support one-way (downlink only) radio science experiments, especially in radio occultation experiments \cite{3GM-2017AGU, Iess+2023SSR}. The spacecraft's antennas are comprised of a dual-band X and Ka steerable Medium-Gain Antenna (MGA), two X-band Low-Gain Antennas (LGA), and a fixed High-Gain Antenna (HGA). The X- and Ka-band capable fixed 2.54~m HGA is to be used to deliver the data downlink. The baseline assumption is that during the early and late segments of each ground station pass, when the ground station antenna elevation is low, housekeeping data will be delivered at X-band. Segments with higher ground station antenna elevations will be used for delivering the scientific data in the Ka-band. 
 
 PRIDE will conduct its measurements in various configurations of the JUICE mission radio systems, including their onboard and Earth-based segments. As demonstrated in the experiments with ESA's Mars Express \cite{Duev+2016, Bocanegra+2018} and ESA's Venus Express \cite{Bocanegra+2019}, PRIDE can operate in the one-way (open loop, downlink only) configuration or the so called ``three-way'' configuration. In the one-way configuration, it is highly beneficial to operate with the downlink signal supported by the onboard oscillator of high stability, like USO. This regime has been demonstrated in the VLBI observations of the Huygens Probe during its descent on Titan \cite{SVP+2004, Lebreton+2005}. The ``three-way'' configuration includes a two-way closed-loop radio link (up- and downlink from/to an Earth-based tracking station) and ``eavesdropping'' of the downlink by a third station, a radio telescope or, more often, an array of radio telescopes involved in PRIDE observations. The analysis of the error budget for various PRIDE regimes is given in \cite{Bocanegra+2018}. Operational arrangements of PRIDE are discussed in \cite{Vidhya+2023ASR}.

PRIDE is capable of conducting its observations with any radio signal emitted by spacecraft during telemetry, tracking or radio science operations. The main operational interface for conducting PRIDE measurements occurs between the JUICE mission and the PRIDE network of ground-based assets (see section~\ref{ss:PRIDE-network}). Due to the nature of its measurements, PRIDE is synergistic to the JUICE radio science instrument 3GM \citep{iess20133gm, Iess+2023SSR}. It is important to underline that PRIDE-JUICE does not require any specific on-board instrumentation beyond those devices which are available on board the mission spacecraft independently of PRIDE, essentially the radio communication instrumentation. Nominally, PRIDE is an experiment exploiting X-band (8.4~GHz) JUICE downlink signal. However, as demonstrated with test observations of the ESA's BepiColombo spacecraft, PRIDE is able to operate with the Ka-band (32~GHz) downlink signal too (see sub-section~\ref{ss:PRIDE_Dop}. PRIDE is sufficiently flexible to operate in a dual-frequency X/Ka mode, if requested to do so in the interests of specific scientific opportunities or operational needs.

To implement the PRIDE organisational and observational plans properly and consistently, the experiment's team must interact with the JUICE science working group, Mission Operations Centre (MOC) and Science Operations Centre (SOC). The MOC and the SOC of the ESA ground segment are in charge of operating the JUICE mission. While SOC is in charge of organising, preparing, and delivering the science operation requests to the MOC and coordinating the distribution of the data collected from the MOC. MOC is responsible for ground segment development and spacecraft operations. It is anticipated that the major load of supporting JUICE science operations will be taken by the ESA's Estrack ground station Malargue (Argentina) in both the X- and Ka-bands. The spacecraft HGA or MGA will be pointed to Earth about 8~hours every day during telemetry passes and the MGA will remain pointed toward Earth during each flyby.

PRIDE is able to operate in the one-way or three-way link mode during these communication time slots. Thus, the main operational interface between the PRIDE and the mission will contain information on schedules of the spacecraft and mission ground stations. Based on this information, PRIDE will schedule ground-based radio astronomy resources (subsection~\ref{ss:PRIDE-network}) as described in section~\ref{s:schd}.

\subsection{Earth-based segment of PRIDE}
\label{ss:PRIDE-network}

\begin{figure}[!hbt]
    \centering
     \includegraphics[width=0.77\columnwidth]{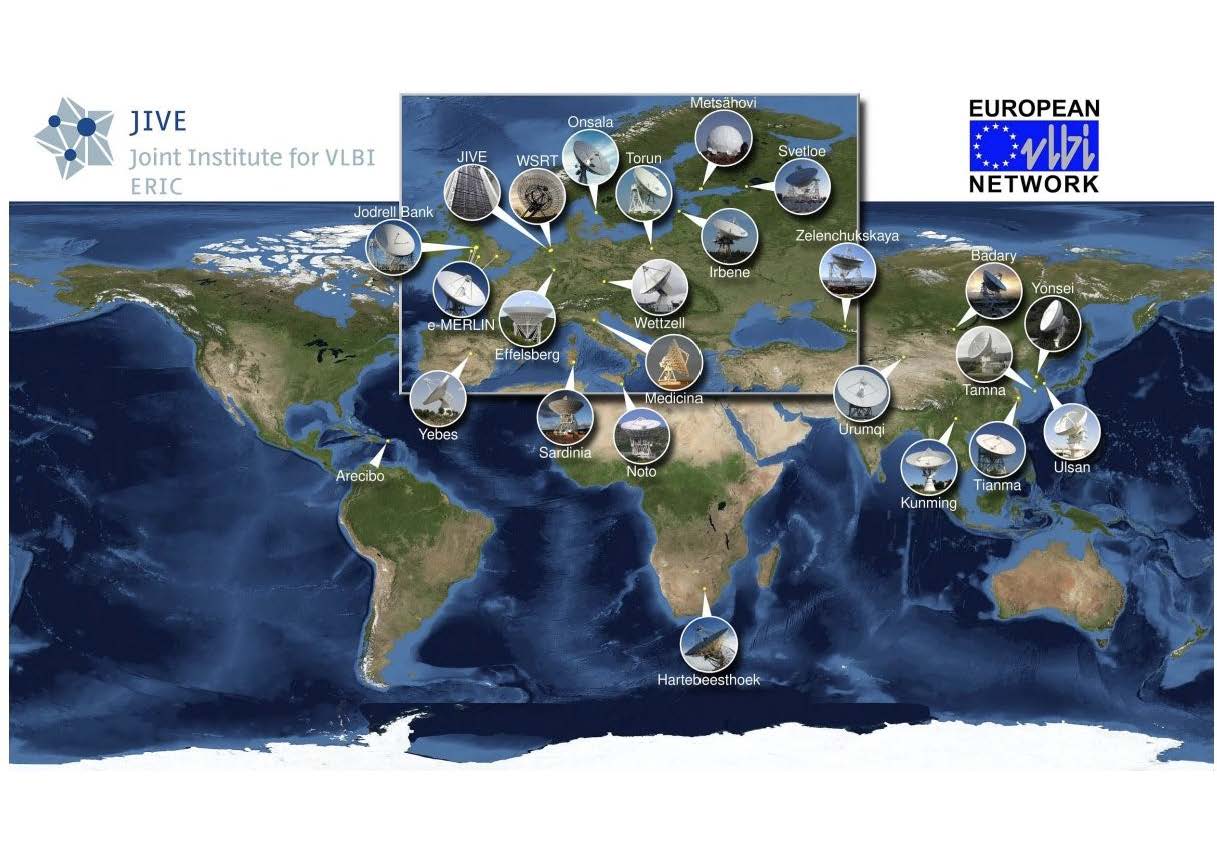}
     \includegraphics[width=0.77\columnwidth]{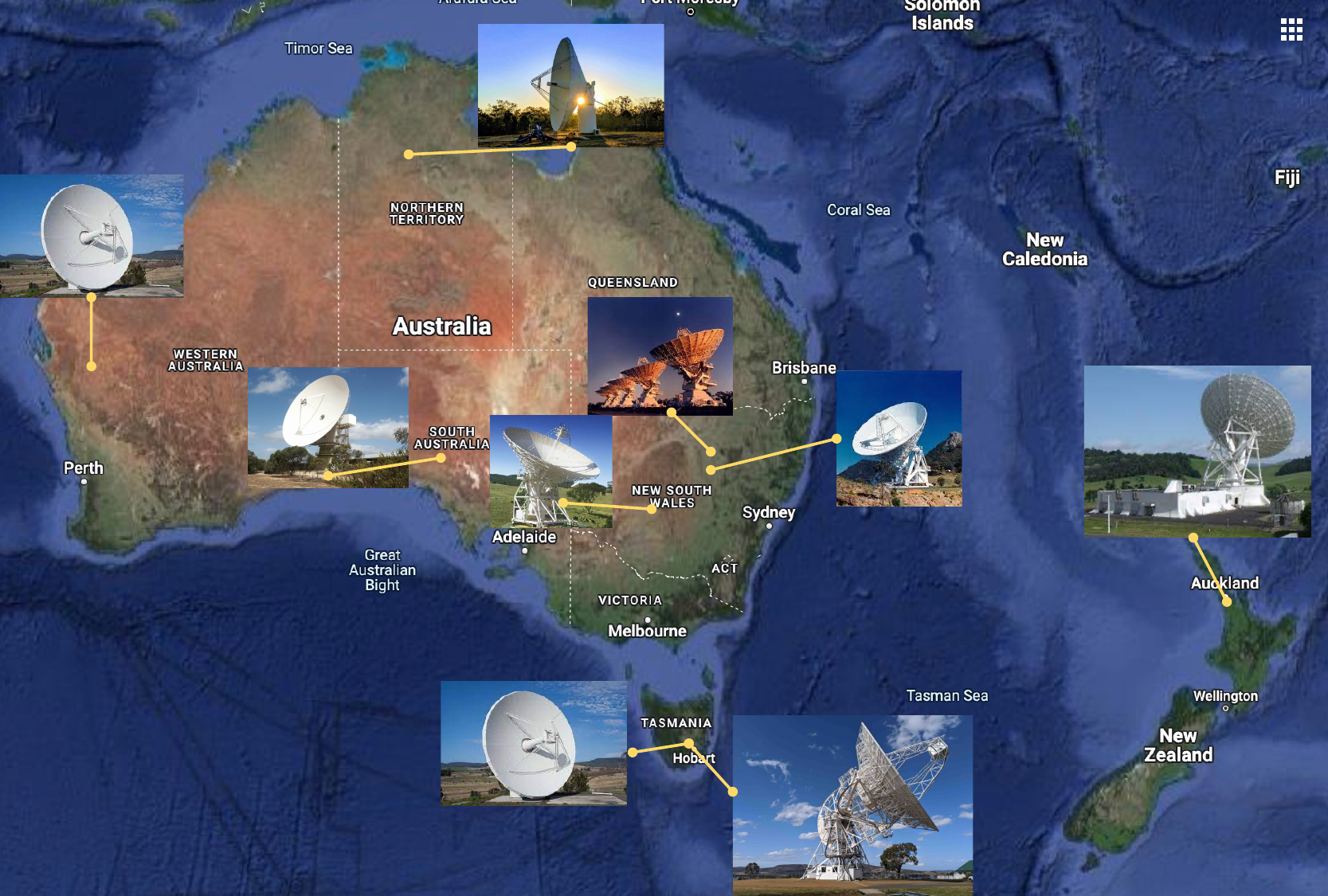}
    \caption{Geographical distribution of some VLBI radio telescopes capable to conduct PRIDE observations. Top: the European VLBI Network (EVN), credit JIVE. Bottom: the Long Baseline Array (LBA) and radio telescopes of the University of Tasmania (UTAS), credit UTAS.}
    \label{fig:EVN-map}
\end{figure}

The Earth-based segment of PRIDE consists of networks of VLBI-equipped radio telescopes spread over the globe (Fig.~\ref{fig:EVN-map}), data transfer infrastructures, and data processing facilities.

A VLBI array is a network of physically disconnected radio telescopes that perform scheduled simultaneous observations (see \ref{ss:sched-vlbi}). Table \ref{tab:vlbi-arrays} lists operational VLBI networks as of the time of this writing in June 2023. The majority of VLBI arrays operate most of time under the open-sky policies, \emph{i.e.}, they provide observing resources solely on the basis of peer-reviewed proposals. Under this access method, observing time is allocated based on the scientific merits of the proposed experiments. However, a number of radio observatories commit their observing resources to specific scientific programs, including VLBI tracking support to planetary and space science missions. PRIDE observations have been already identified as a scientifically meaningful VLBI application that has innovative technical and scientific results. Arrangements for PRIDE observations as a part of the JUICE mission have been warranted under both, open sky access and direct observing commitments mechanisms. In addition to the radio telescopes involved in the VLBI networks and arrays listed in Table~\ref{tab:vlbi-arrays}, about a dozen of antennas around the world are nominally unaffiliated with VLBI networks or operate primarily as deep space communication facilities, like the NASA's Deep Space Network (DSN), but are fully equipped for VLBI observations. All together, at present, about 60 antennas are capable to conduct VLBI observations. This number is likely to increase during the decade of the JUICE in-flight operations. All VLBI telescopes are equipped with highly stable frequency standards (typically -- hydrogen maser oscillators). A majority of the VLBI telescopes are equipped with receivers able to observe at the nominal PRIDE frequency of 8.4~GHz (the downlink communication X-band). Some of the currently operational antennas are also able to observe at the Ka-band (32~GHz). 

\begin{table}[h]
\caption{The world VLBI arrays. }
\begin{tabular}{@{}lcc@{}}
\toprule
\textbf{VLBI Array} & \textbf{No. of telescopes}\footnotemark[1] & \textbf{Reference}\footnotemark[2] \\
\midrule
European VLBI Network (EVN) & 22 & [a] \\
Very Long Baseline Array (VLBA) & 10 & [b] \\
Japanese VLBI Network (JVN) & 9 & [c] \\
Chinese VLBI Network (CVN) &  $>$5  & [d] \\
Korean VLBI Network (KVN) & 3 & [e] \\
East-Asia VLBI Network (EAVN) &  15  & [f] \\
Australian Long Baseline Array (LBA) & 5 & [g] \\
University of Tasmania Array & 5 & [h,i] \\
\botrule
\end{tabular}
\footnotetext[1]{Some telescopes might participate in more than one network.}
\footnotetext[2]{All web references have been verified on 2023.06.02.}
\begin{tabbing}
[x] \= *************************************** \= \kill
[a] \> \url{https://www.evlbi.org} \\\relax
[b] \> \url{https://science.nrao.edu/facilities/vlba} \\\relax
[c] \> \url{https://www.miz.nao.ac.jp/en/content/project/japanese-vlbi-network.html} \\\relax
[d] \> \url{http://astro.sci.yamaguchi-u.ac.jp/eavn/aboutcvn.html} \\\relax
[e] \> \url{https://radio.kasi.re.kr/kvn/main.php} \\\relax
[f] \> \url{https://radio.kasi.re.kr/eavn/main.php} \\\relax
[g] \> \url{https://www.atnf.csiro.au/vlbi/overview/index.html} \\\relax
[h] \> \url{https://space.phys.utas.edu.au} \\\relax
[i] \> \url{https://auscope.phys.utas.edu.au/} \\\relax
\end{tabbing}
\label{tab:vlbi-arrays}
\end{table}

The data collected at the radio telescopes containing the spacecraft signal are either streamed using high-speed data links, or physically shipped, from the telescopes to the central data processing (correlation) facility at JIVE in the Netherlands. The corresponding logistical infrastructure is not mission-specific and is being used routinely for all types of VLBI observations. 

PRIDE data recorded at different telescopes are processed jointly in the correlation processing cluster (correlator). The SFXC correlator (Software FX-kind Correlator, \citep{2015ExA....39..259K}), the main data processing correlator of EVN is designated for processing PRIDE-JUICE observations. This process consists of matching the signal arrival times and Doppler shifts at different telescopes to align them properly in delay and delay-rate space. The SFXC correlator at JIVE has demonstrated its ability to process spacecraft VLBI tracking data in the near-field VLBI mode (see \cite{Duev+2012, Duev+2016} and further description in section \ref{s:PRIDE-proc}). The output of the correlator is written into Flexible Image Transport System (FITS) files. The data contained in these files should be seen as the raw data of PRIDE VLBI measurements. Following the established practice of EVN operations, these files are stored at the central archive maintained at JIVE. 

\subsection{PRIDE data handling and processing hardware}
\label{ss:PRIDE-data_pf}

PRIDE is an off-spring of mainstream VLBI developments. The latter is primarily aimed at investigation of compact structures of galactic and extragalactic radio sources (astrophysical VLBI), precise determination of their celestial coordinates (astrometric VLBI) and precise measurements of the radio telescope terrestrial coordinates and Earth rotation parameters (geodetic VLBI). All these developments are chartered for the period 2020--2030 and beyond \cite{EVN2030arXiv}, de facto covering the operational lifetime of the JUICE mission. It is also expected that during the cruise phase of the JUICE mission the next generation radio telescope, the Square Kilometre Array (SKA) will begin its operations. The SKA will operate in VLBI modes of various ``flavours'' \cite{SKA-VLBI-2015}, supported by the required data handling and processing instrumentation. VLBI data processing involves massive computing at the correlation and post-correlation phases. PRIDE data processing, including the post-correlation phase requires the same hardware as other applications of VLBI. This hardware is to great extent subjected to Moore's law (doubling of digital electronics capability every two years, \cite{Moore1965cramming,Birnbaum+2000}). Even the most pessimistic prognosis of the remaining latency of this law through the 2020s \cite{Rotman2020MIT} allows us to safely assume that the needs of VLBI data processing can rely on the so far outpacing development of digital electronics. Since the operational duration of the JUICE mission is about 12 years, it is reasonable to expect that at least several ``doubling'' Moore's cycles will result in significant modifications of the mainstay computational hardware used in VLBI. Furthermore, new developments in processing power of Graphic Processing Units (GPU) are moving the industry towards dedicated GPU clusters that are more efficient in correlating \citep{yu+al+2023}. Thus, it is safe to assume that PRIDE-JUICE needs in data handling and processing hardware will be well within the capabilities of the prospective VLBI systems \cite{EVN2030arXiv,SKA-VLBI-2015} consistent with the expected progress of digital electronics during the JUICE operational lifetime.

\section{PRIDE algorithms and data processing}
\label{s:PRIDE-proc}

A set of dedicated software tools has been developed for data processing of PRIDE observations of spacecraft targets in the near-field. The software is able to process narrow-band signals generated by spacecraft, and broadband signals natural celestial radio sources. Within the framework of PRIDE,  the latter is required for calibration and phase-referencing (see section~\ref{s:ph-ref-catalog}) purposes. These software tools are described in the following sub-sections.


\subsection{Spacecraft carrier signal extraction and Doppler measurements}
\label{ss:PRIDE_Dop}

Data processing of the narrow band spacecraft radio signal is conducted with the Spacecraft Doppler tracking software (SDtracker) developed by the PRIDE group. The methodology of the processing is presented in \cite{Duev+2012,Bocanegra+2018} and its algorithmic implementation in~\citep{Molera+2021}. The software is published under MIT license and accessible via a public \textit{git} repository\footnote{\url{https://gitlab.com/gofrito/sctracker/}, accessed on 2023.06.20.}.

\begin{figure}[!htb]
    \centering
     \includegraphics[scale=0.60]{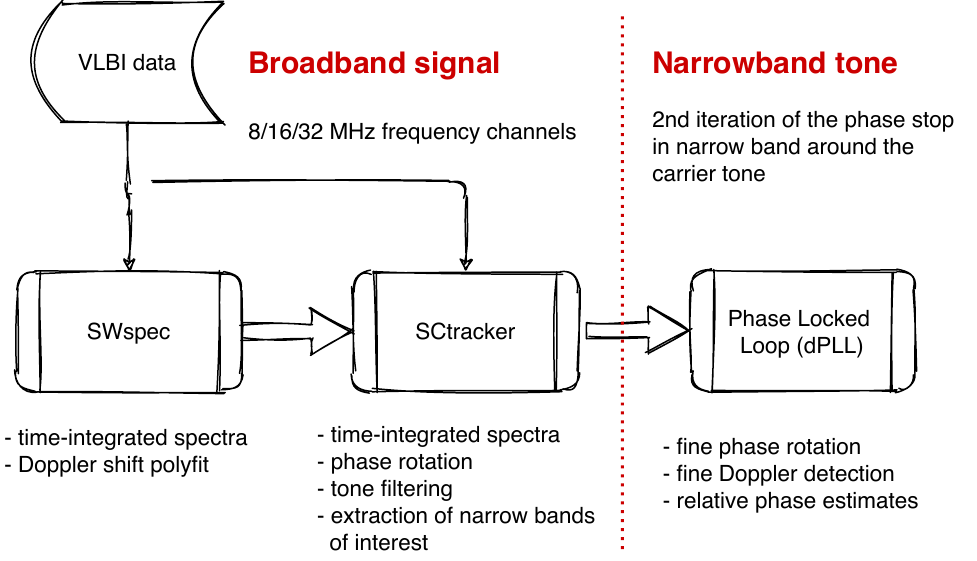}
    \caption{The spacecraft signal observed by a VLBI-equipped radio telescope is processed by three software components: SWspec, SCtracker, and dPLL. The main data products of SDtracker are topocentric frequency detections and the residual phase of the spacecraft carrier and sub-tones~\cite{Molera+2021}.}
    \label{fig:dsp}
\end{figure}

SDtracker consists of three distinct open source packages that process data acquired by a VLBI-equipped radio telescope to generate several data products. The VLBI data format differs from that of nominal deep space tracking and communication systems, such as NASA DSN and ESA Estrack. While the latter provide frequency measurements with respect to the initial transmitted tone \cite{CCSDS401.0-B-29}, radio astronomical systems measure the topocentric frequency and residual phase of the spacecraft carrier radio signal recorded in a broader bandwidth (typically from several hundred MHz to GHz). 

The software package consists of the following three components (Fig.~\ref{fig:dsp}):
\begin{itemize}
 \item Software spectrometer (SWspec)
 \item Multi-tone spacecraft tracker (SCtracker)
 \item Digital Phase Locked Loop (dPLL)
\end{itemize}

\begin{figure}[!htb]
    \centering
    \includegraphics[scale=0.50]{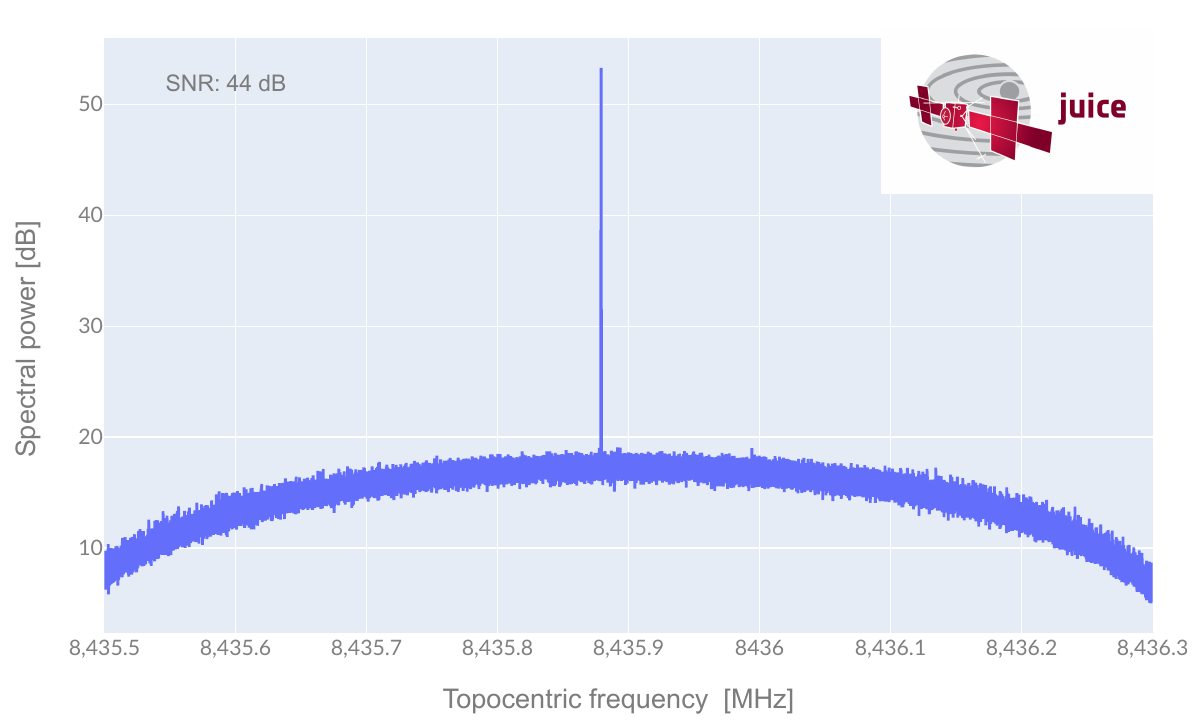}
    \caption{Spectral power in a 1~MHz bandwidth of the ESA's JUICE carrier signal at X-band as received by the 12-m antenna at Hobart (Tasmania, Australia) on 2023 May 20.}
    \label{fig:toneХ}
\end{figure}

SWspec computes a time-series of the signal power spectra in the whole available band, with a selectable spectral resolution from several kHz to sub-Hz. SWspec provides the initial detection and estimation of the Doppler shift and its variation with time. In Fig.~\ref{fig:toneХ} presents the carrier tone of the JUICE spacecraft at X-band ($8435.98$\,MHz) in a bandwidth of $1$\,MHz as observed by the VLBI radio telescope in Hobart (Tasmania, Australia) on 2023 May 20.

\begin{figure}[!htb]
    \centering
     \includegraphics[scale=0.42]{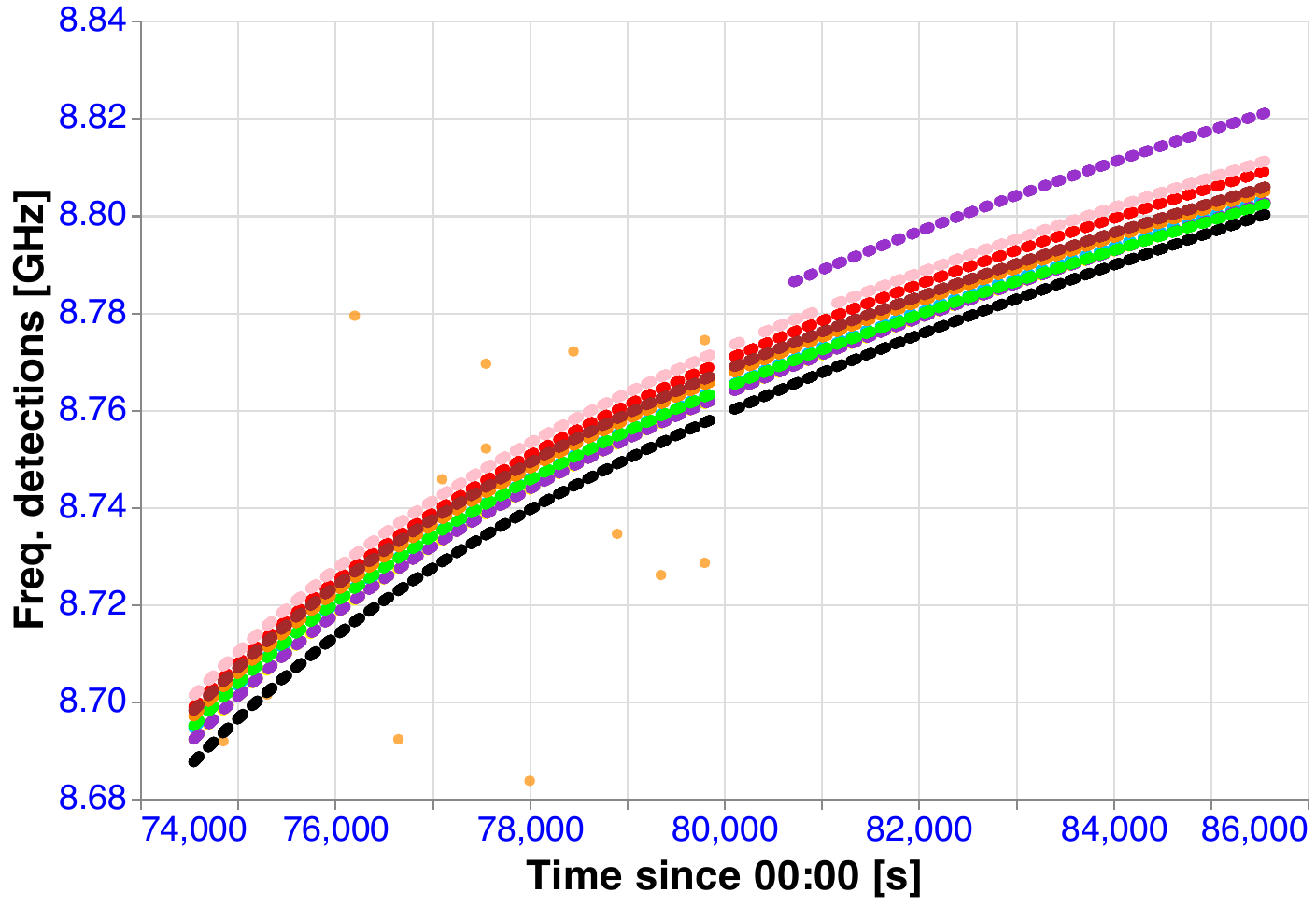}
    \caption{Apparent topocentric frequency detections of the Mars Express signal observed with multiple radio telescopes in a phase-referencing session on 2020 June 21. The observations were conducted at X-band and spanned over three hours. The data were obtained by ten radio telescopes, represented on the plot by different colours.}
    \label{fig:fdets}
\end{figure}

The data products generated consist of a set of frequency and phase polynomial coefficients of a selectable order (usually from 4 to 6), and the revision 0 (or raw) of the topocentric frequency detections. The phase polynomial coefficients are stored on a disk for further data processing in SCtracker. The apparent topocentric frequency detections observed with multiple radio telescopes using the phase-referencing technique can be seen in Fig.~\ref{fig:fdets}.

SCtracker uses the previously estimated phase polynomial coefficients to effectively compensate for the Doppler shift to a first degree of accuracy.
It conducts the initial phase-stop by doing a phase-rotation of the entire recorded bandwidth. The desired carrier frequency is selected to perform signal filtering, to extract a narrow band around the carrier, or multiple tone, with the selected bandwidth, and to detect the relative phase of the tone. SCtracker is capable of simultaneously track the spacecraft carrier signal, its subcarriers and all the ranging tones. SCtracker provides the topocentric frequency detection revision~1.

\begin{figure}[!htb]
    \centering
    \includegraphics[width=0.84\columnwidth]{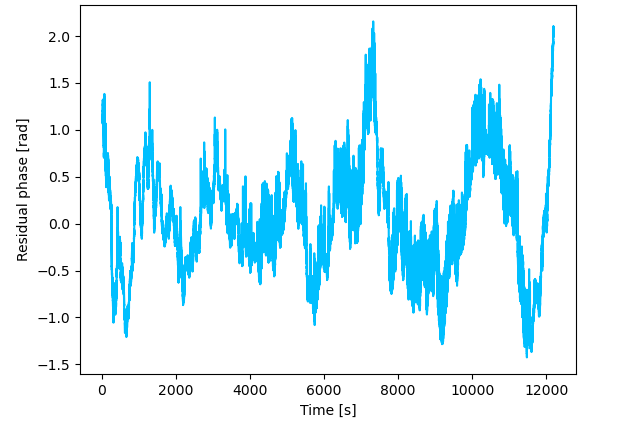}
    \caption{Residual phases of the JUICE carrier signal at X-band as detected by the 12-m radio telescope at Hobart (Tasmania, Australia) on 2023 May 20. The residual phase is extracted relative to the Hydrogen-maser frequency standard at the radio telescope site.}
    \label{fig:phases}
\end{figure}

dPLL provides a finer detection of the carrier signal in an ultra-narrow band, and it calculates the topocentric frequency measurements and the residual phase with respect to the station clock. It repeats the steps of SWspec and SCtracker on the filtered narrow band of the spacecraft signal at a much higher spectral resolution. The output of SDtracker provides the power spectrum profile, topocentric frequency detections with a sub-mHz resolution (revisions 2 and 3), the residual phases of the carrier and the sub-carrier's tone with high spectral resolution with time stamps of the station's clock. The residual phase for JUICE signal is shown in Fig.~\ref{fig:phases}.

\begin{figure}[!htb]
    \centering
    \includegraphics[width=0.84\columnwidth]{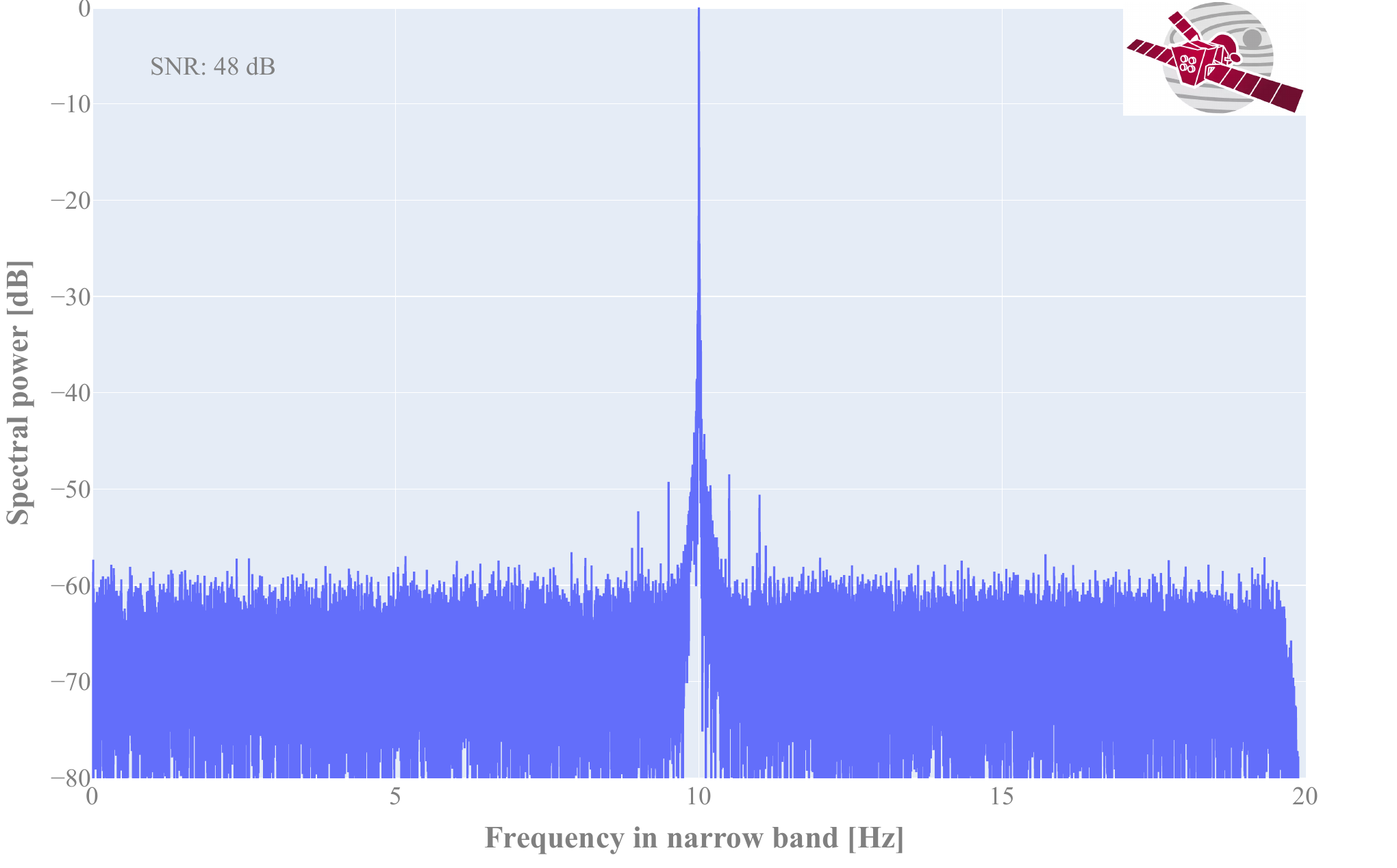}
    \caption{Spectral power in a 20 Hz narrow band of the ESA's BepiColombo carrier and sync words signals at Ka-band as received by the 24-m antenna at Mopra (NSW, Australia) on 2022 April 11.}
    \label{fig:toneKa}
\end{figure}

The software described above has been tested with radio signals from various spacecraft and various communication bands. Fig.~\ref{fig:toneKa} presents the Ka-band carrier tone (32200~MHz) within the band of 20~Hz of the ESA's BepiColombo mission observed by the VLBI radio telescope Mopra (New South Wales, Australia) on 2022 Apr 11. The observational data from this observation were processed with the software described above.

\subsection{Near-field VLBI measurements}
\label{ss:nfVLBI}

\begin{figure}[h!]
    \centering
     \includegraphics[width=0.84\columnwidth]{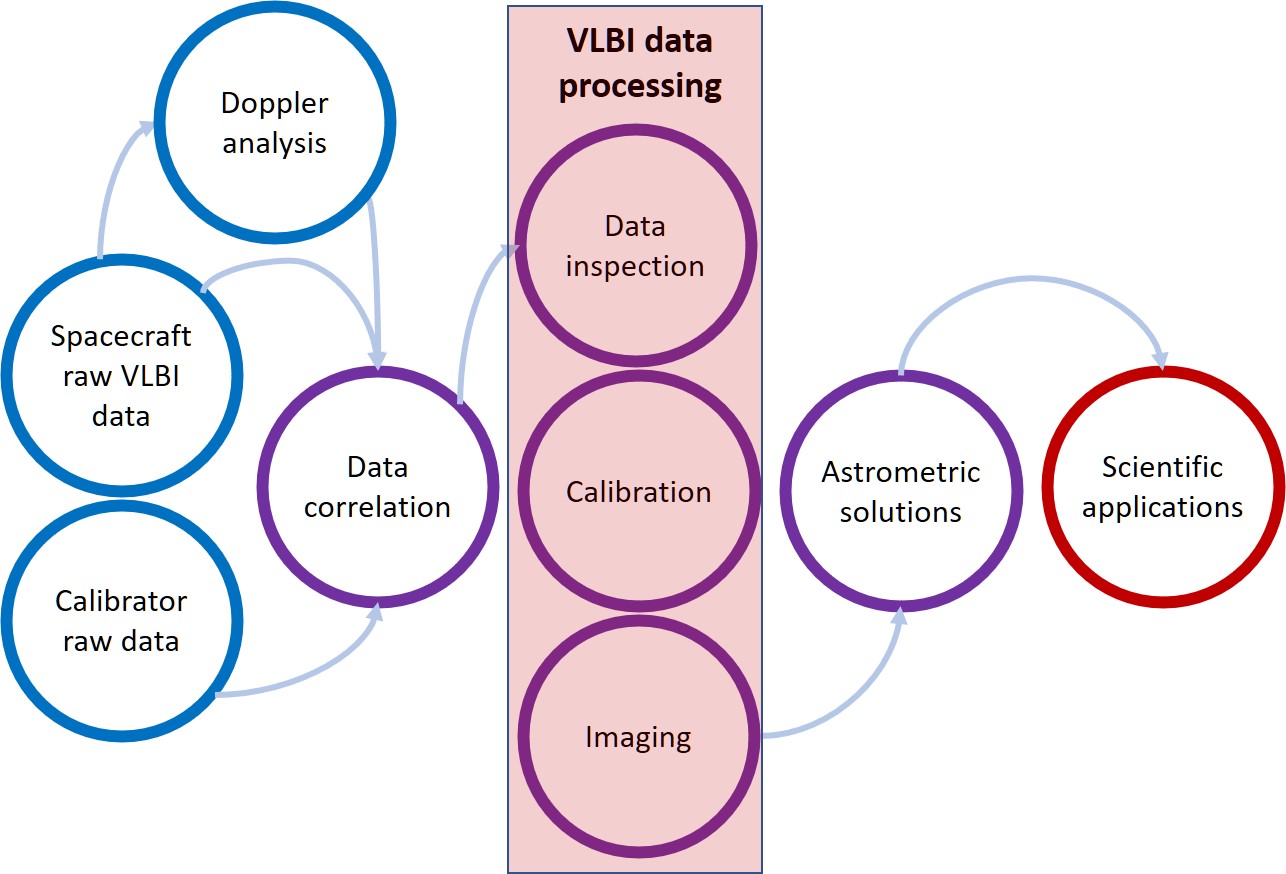}
    \caption{Flowchart of the VLBI data processing from the raw data that contain the spacecraft and calibrator signal to the experiment science products.}
    \label{fig:VLBi-data-processing}
\end{figure}

In many ways, PRIDE observations do not differ from standard astronomical VLBI observations, which typically involve classic VLBI targets like natural radio sources from nearby objects in the Galaxy to objects at cosmological distances. Fig.~\ref{fig:VLBi-data-processing} illustrates major steps in data processing of VLBI tracking as applicable to spacecraft PRIDE tracking. Spacecraft VLBI tracking requires an adaptation of the phase-referencing VLBI technique for a near-field position of a target spacecraft and estimating its celestial position by observing a known nearby source, called a phase-reference calibrator. This technique requires alternated pointing of the telescopes to the target, such as a spacecraft, and the calibrator, unless they are close enough in the sky to be within the primary beam of a telescope involved in the observation, which is termed ``in-beam phase referencing'' and provides the most accurate measurements. For phase referencing, the target and calibrator sources should generally be within a few degrees apart \cite{Ros+1999AA}. 

As discussed in Section \ref{s:intro}, for sources within the Solar System, it is necessary to account for the delay in the signal's arrival time at Earth-based telescopes that is corrected for the near-field geometry. The PRIDE team has formalised the method for calculating near-field delays and determining spacecraft state vectors with sub-milliarcsecond accuracy in \citep{Duev+2012}.

The main difference between correlating standard VLBI observations and PRIDE data lies in the need for an ad hoc "near-field" delay model. PRIDE data are correlated at JIVE using the SFXC software correlator \citep{2015ExA....39..259K}, which is based on the original design developed for VLBI tracking of the Huygens probe \citep{SVP+2004} and supports both far-field and near-field delay models.

The VLBI delay model is formulated in the Barycentric Celestial Reference System (BCRS), i.e. the delay is computed in the time scale of the BCRS and is subsequently transformed to the time-scale used for timing the measurements on Earth -- to be later used for observational data processing. This necessitates a series of coordinate and time-scale transformations to reduce the observer and target space-time coordinates needed for delay computation into a common frame of reference \citep{soffel2003iau}. 

Station coordinates in the geocentric terrestrial reference system (realised by the International Terrestrial Reference Frame, ITRF) are reduced to the epoch of observation, then transformed to the Geocentric Celestial Reference System (GCRS) following the latest recommendations of the international Earth rotation service. These transformations account for the motion of the celestial pole in the GCRS, Earth's rotation around the pole-associated axis, and polar motion. Station positions in GCRS are corrected for geophysical effects, including displacements caused by solid Earth tides due to the direct effect of external tide-generating potential, ocean tidal loading, diurnal and semidiurnal atmospheric pressure loading, and centrifugal perturbations from pole tide-driven Earth rotation variations \citep{petit2010}. Ultimately, the Lorentz transformation is applied to the corrected station positions in GCRS to transform them to BCRS.

The time-scale used for measurement timing at stations is Coordinated Universal Time (UTC), while ephemerides of planetary spacecraft and solar system bodies typically use Barycentric Dynamical Time (TDB). The UTC-to-TDB transformation involves several steps and considers conventions, such as leap seconds, and relativistic effects, like gravitational potential from Earth and the observatory's diurnal speed effects on terrestrial clocks.

Signal delay, defined as the difference between two light travel times from the spacecraft to two stations forming a VLBI baseline, is computed using an iterative procedure in the barycentric TDB-frame and then transformed into the geocentric frame. The resulting signal delays are corrected for instrumental and propagation factors, including tropospheric and ionospheric effects. Formal mathematical descriptions of these calculations are given in \citep{Duev+2012}.


With VLBI delays determined, the correlation of both the natural phase calibrator, assumed to be at an infinite distance from the observer, and the near-field spacecraft data can proceed. A two-way Doppler phase correction (obtained with the SCTrack software, as referenced in Section \ref{ss:PRIDE_Dop}) is applied to the spacecraft data to prevent frequency smearing. The correlation results are saved in standard file formats for data analysis with common radio astronomical software packages. For a detailed explanation of the spacecraft data correlation process, see \citep{Duev+2016}.

\begin{figure}[h!]
    \centering
    \includegraphics[width=0.94\columnwidth]{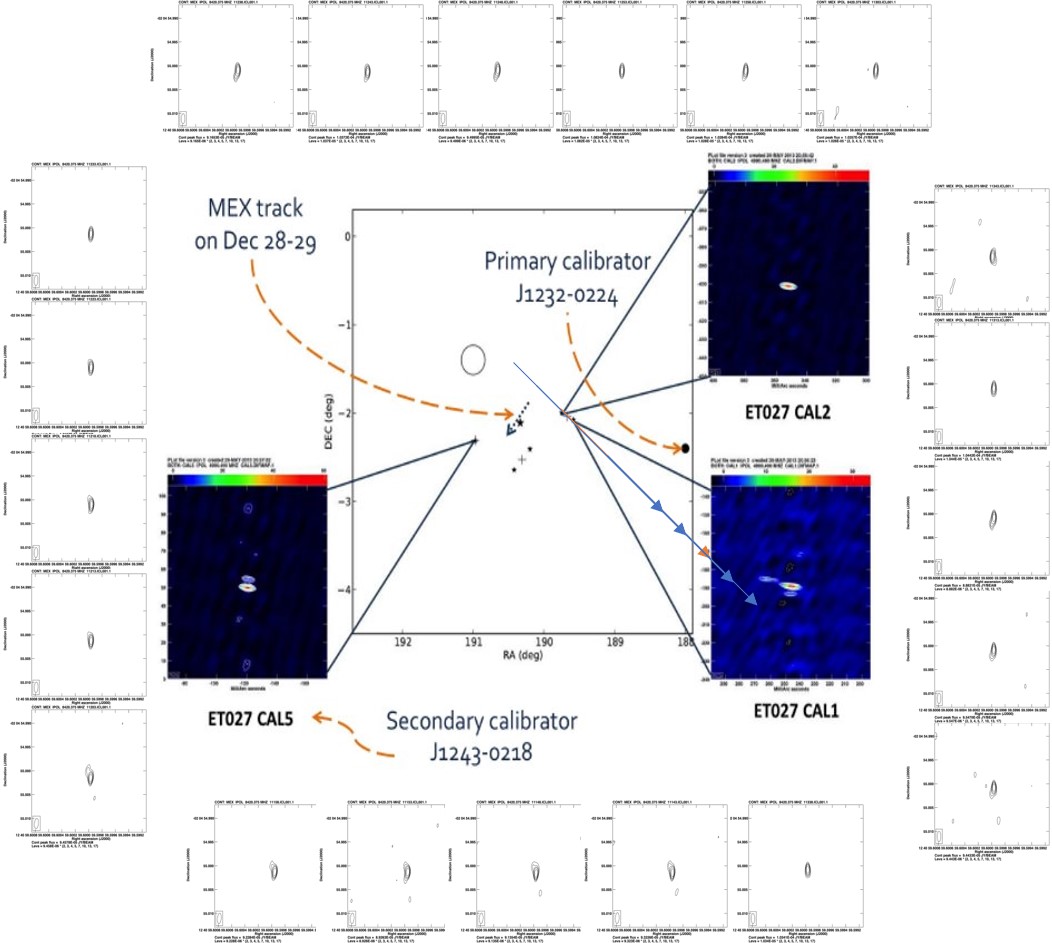}
    \caption{\textit{Inner part:} A celestial finding chart of the PRIDE observation of the ESA's Mars Express spacecraft during its flyby of Phobos on 2013.12.29 (Global VLBI project GR035). The color insets show VLBI images of the phase-referencing calibrators. \textit{Outer part:} Grey-scale images of the MEX spacecraft in 2-minutes intervals. Adopted from \cite{Duev+2016}. }
    \label{fig:mex-phobos}
\end{figure}

The post-correlation data reduction is performed using VLBI standard software packages, such as AIPS \footnote{\url{http://www.aips.nrao.edu/}, accessed on 2023.06.20} (Astronomical Image Processing System)  and CASA\footnote{\url{https://casa.nrao.edu/}, accessed on 2023.06.20.} (Common Astronomy Software Applications). The first step is a preliminary data inspection and editing. Then the initial calibration is applied, to correct for the irregular shape of the bandpass in the receivers channels and to scale the amplitudes measured by each telescope for representing them in physical units of flux density\footnote{The conventional unit of flux density in radio astronomy is called jansky (Jy); $1$\,Jy\,$= 10^{-26}$\,W\,m$^{-2}$\,Hz$^{-1}$.}. After this, instrumental and atmospheric effects are corrected by using the phase calibrator measurements to calibrate the target source. This process uses the standard VLBI algorithm of fringe-fitting to calculate the delays and rates of the phase referencing calibrator, which are later applied to the spacecraft. Once the calibration process is completed, a map of the calibrator and the spacecraft can be produced. Usually this is done using the {\tt CLEAN} algorithm (\citep{1974A&AS...15..417H}). The position measured on the map in Right Ascension and Declination provides the actual coordinates of the spacecraft. The process can be repeated for each observing interval.

\begin{figure}[h!]
    \centering
    \includegraphics[width=0.99\columnwidth]{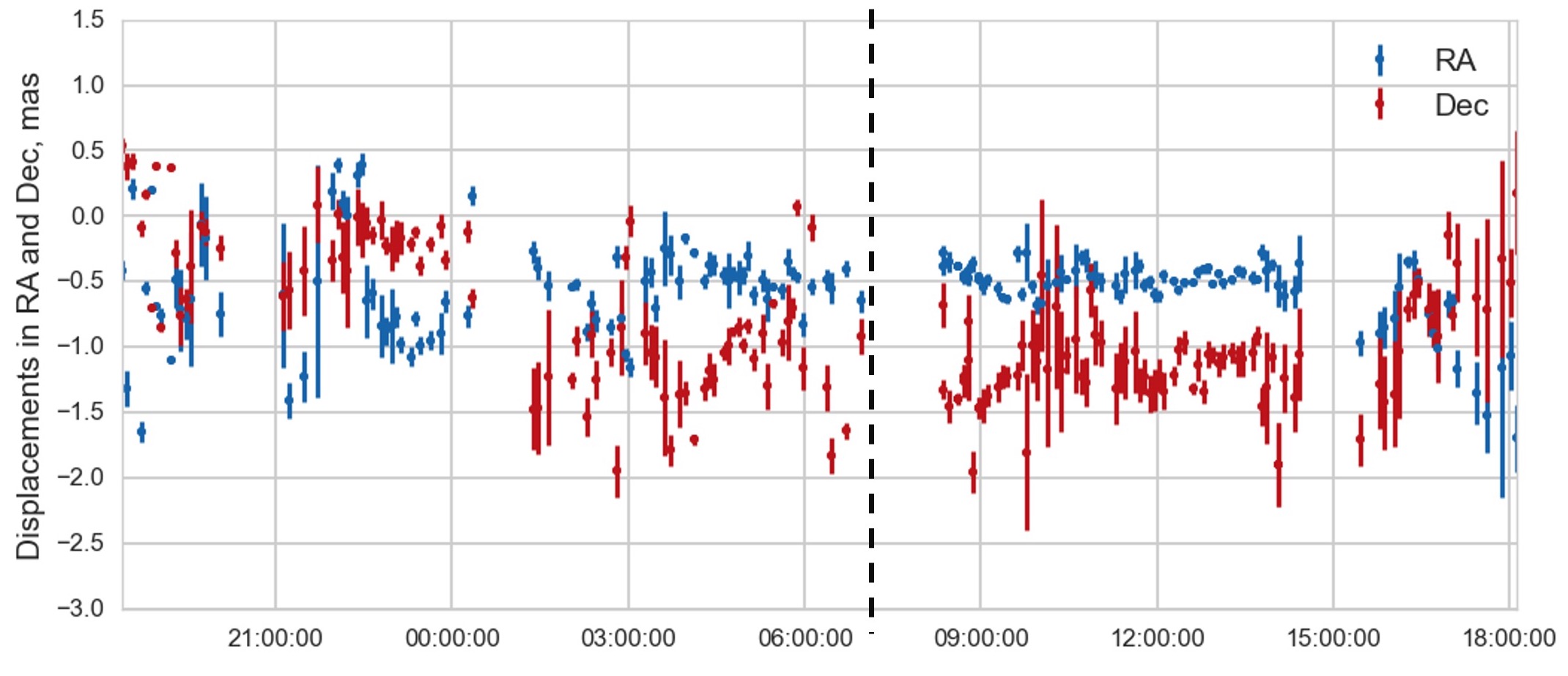}
    \caption{Displacements from the a priori lateral position of MEX as a function of time, measured using the PRIDE observations during the Phobos flyby on 2013.12.28-29. Displacements in Right Ascension (mas) are shown in blue, in Declination (mas) in red, 2-min integration time. The closest to Phobos approach on 2013.12.29 at about 07:09 UTC is indicated by the vertical black dashed line. See \cite{Duev+2016} for further details.}
    \label{fig:gr035eph}
\end{figure}

PRIDE observations might be of special value during JUICE spacecraft flybys of various celestial objects. Fig.~\ref{fig:mex-phobos} shows the results of the PRIDE observations of the flyby of Phobos by the ESA's Mars Express (Global VLBI project GR035). An image of the spacecraft was obtained every two minutes for 26 consecutive hours. This allowed us to follow MEX during three complete orbits around Mars, including the one (the middle one) containing the flyby itself. More than 30 radio telescopes distributed around the world participated in these observations. The experiment resulted in the improvement of the knowledge of the MEX ephemeris comparing the best a priori values by $\sim0.5$\,mas in right ascension and $\sim1.0$\,mas in declination, Fig.~\ref{fig:gr035eph}.

\begin{figure}[h!]
    \centering
    \includegraphics[width=0.7\columnwidth]{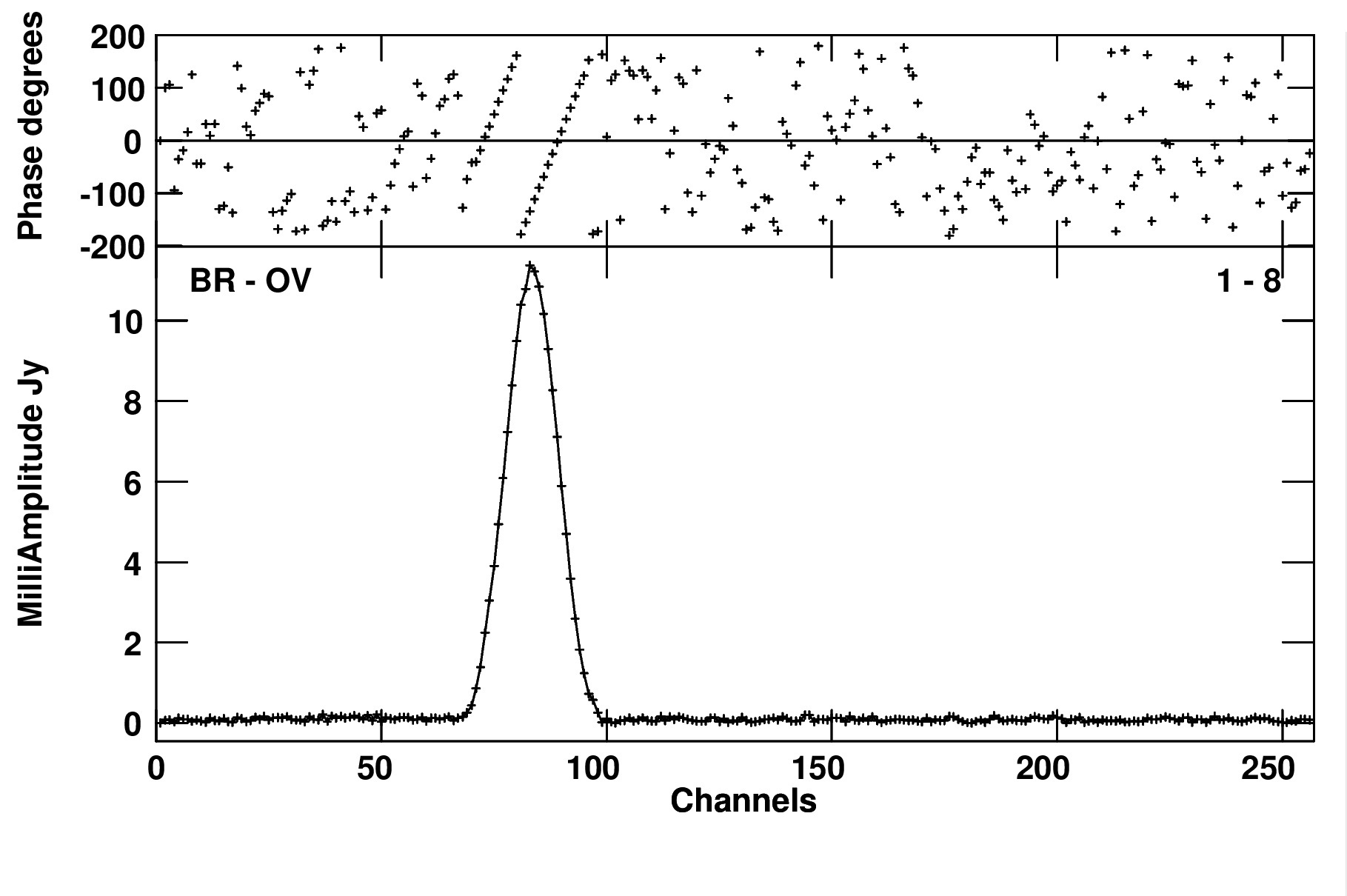}
    \caption{VLBI detection on a VLBA baseline of the NASA's Mars Reconnaissance Orbiter while pointing the VLBA antennas at Mars (VLBA experiment BC246A). \emph{Bottom:} The amplitude of the cross-correlated response of the spacecraft signal plotted versus the frequency channels. \emph{Top:} The phases of the cross-correlated spacecraft's transmission show a systematic slope in the frequency channels 65--100 that indicates an uncorrected delay mostly due to the offset in the position of the spacecraft. Accurate determination of this offset provides precise estimates of the spacecraft position -- the main observable of PRIDE.
    }
    \label{fig:mro}
\end{figure}

Similar to the demonstrated above PRIDE observations of MEX during its Phobos flyby have been carried out in recent years with MEX and other spacecraft. Several spacecraft in orbit around Mars have been targeted to test different PRIDE configurations. In particular, Figure \ref{fig:mro} shows the detection of NASA's MRO spacecraft by the VLBA array on the Brewster and Owens Valley baseline. During these observations, carried out in 2018, we pointed the telescopes in the direction of Mars with scans of two minutes and were able to observe multiple spacecraft at the same time, all within the same primary beam of individual radio telescopes. 
In the optimal case, both target spacecraft and the calibrating source are within the primary beam of the antennas. No switching between the calibrator and the targets is necessary. In section \ref{s:Eu-Clip}, we will discuss how this special configuration can be applied to simultaneous observations of JUICE and Europa Clipper.

\section{Positioning the JUICE spacecraft in the ICRF with PRIDE}
\label{s:ph-ref-catalog}



Phase-referencing near-field VLBI observations contribute to the determination of the state vector of the JUICE spacecraft. This is done by linking the celestial position of the spacecraft to angularly close calibrator sources whose absolute positions are accurately known in the quasi-inertial celestial reference frame defined by the grid of quasars. Thus, the absolute position of the spacecraft will also be determined in the same quasar-based frame through such observations. The availability of suitable (i.e. sufficiently bright and compact) nearby calibrator quasars is therefore essential for the PRIDE success. Below we introduce the current realisation of the ICRF and comment on the possibility of its next realisation by the time of JUICE arrival at Jupiter. Since the ICRF grid is not sufficiently dense for providing close enough calibrators at most of the time along the spacecraft trajectory, we also outline strategies for the densification of the reference source grid, based on available radio source lists and using dedicated astrometric VLBI experiments prior to the PRIDE observations.

\subsection{International Celestial Reference Frame}
\label{ss:icrf-intf}

 The most accurate celestial reference frame available to date is the third realisation of the ICRF  \citep[ICRF3,][]{2020A&A...644A.159C}, which was adopted by the International Astronomical Union (IAU) General Assembly in 2018. As such, it has become the fundamental celestial reference frame in use since 2019 January~1. The ICRF3 includes a total of 4588~sources covering the entire sky, among which 4536 were measured at X-band (from dual-frequency S/X~observations), 824 at K-band and 678 at Ka-band (from dual-frequency X/Ka~observations), 600~sources being common to the three frequencies. The median positional uncertainty in ICRF3 is about 0.1~mas (0.5~nanoradian) in right ascension and 0.2~mas (1~nanoradian) in declination, with a noise floor of 0.03~mas (0.17~nanoradian) in the individual source coordinates. 

A specific feature of ICRF3 is that the astronomical modeling incorporates Galactocentic acceleration to account for the rotational motion of the Solar System around the Galactic center, meaning that the source coordinates are not anymore fixed with time as in the two previous realisations of the ICRF. Instead, they are subject to a dipolar proper motion field of amplitude 0.0058~mas\,yr$^{-1}$, leading to significant positional corrections for observations away from the ICRF3 reference epoch, which was set to 2015.0. In the case of JUICE, those corrections will be on the order of 0.1~mas when the spacecraft is at Jupiter in the time-frame 2030--2035. This is three times above the ICRF3 noise floor, meaning that these corrections will have to be considered.

The most useful ICRF3 positional information for PRIDE-JUICE is the one available at X- and Ka-bands since the mission down-link (and therefore the planned VLBI phase-referencing measurements) is conducted at those two frequency bands. Carrying out phase-referencing near-field VLBI observations implies that a suitable set of calibrators is available alongside the spacecraft trajectory. Since relative position errors scale with the angular separation between the target and the calibrator \citep{2006A&A...452.1099P}, the direction of those calibrators must be as close as possible as that of the spacecraft. As this may not be directly achievable based on the ICRF3 source list (see below), finding new, closer (though likely weaker) calibrators by means of dedicated astrometric VLBI observations will be essential to allow for positioning of the JUICE spacecraft with the highest accuracy. A critical element in the analysis of such observations will be the placing of the new sources onto the ICRF3 grid. Alternatively, the corresponding data, if conducted in global VLBI astrometry mode, may be incorporated into the sets of data considered when building the next ICRF. Assuming the current pace of a new realisation of the ICRF every decade (1997 for ICRF1, 2009 for ICRF2, and 2018 for ICRF3) is maintained, the availability of a successor for the ICRF3 by the time of arrival of the spacecraft at Jupiter can be envisioned with reasonable probability. The newly-identified calibrators would then be directly part of the next ICRF and no further link would be necessary.

\subsection{Densification of the calibrator list along the JUICE trajectory}
\label{ss:dens-ref}

\begin{figure}[!htb]
    \centering
    \includegraphics[width=0.99\columnwidth]{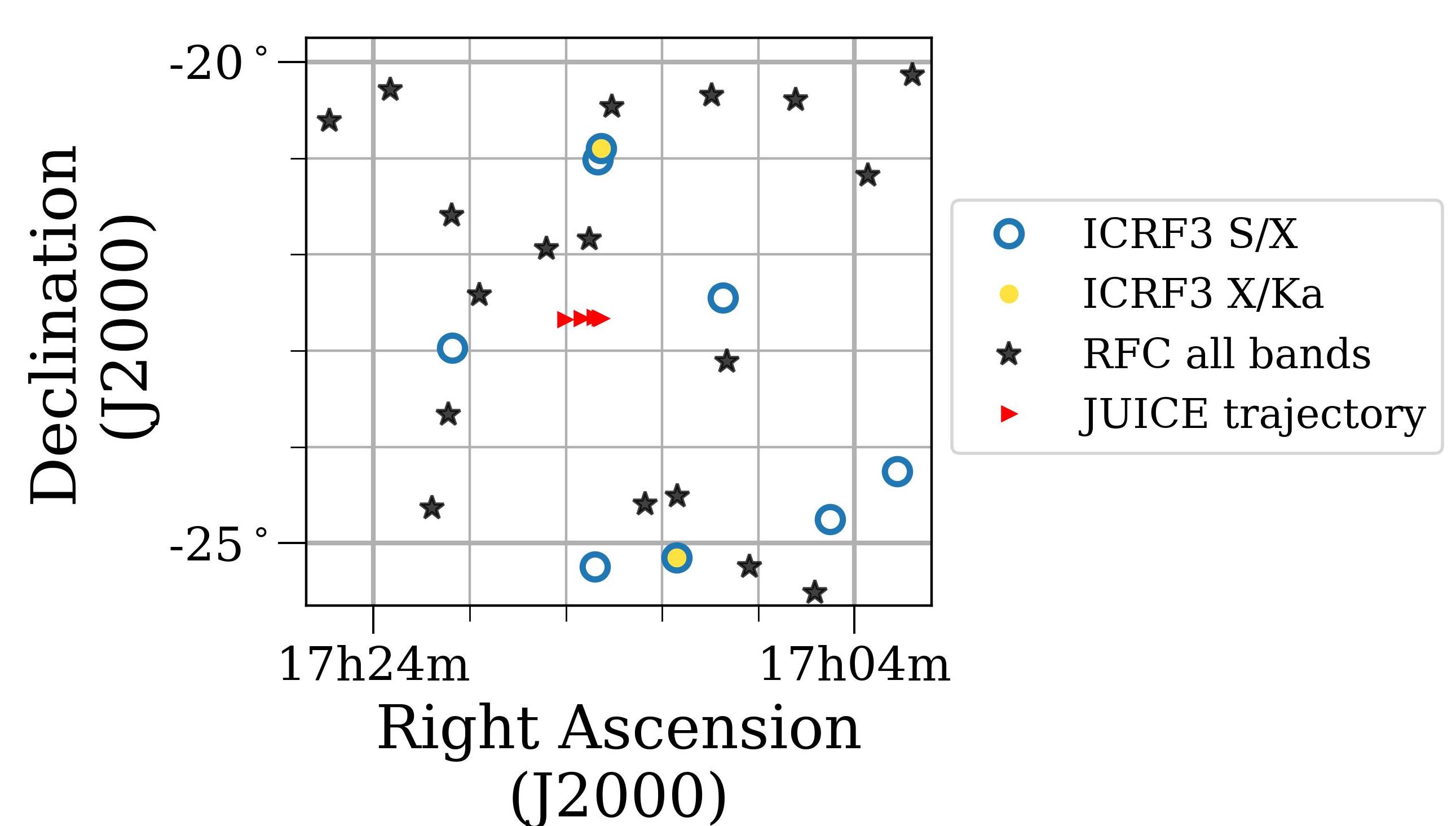}
    \caption{Distribution of the currently known ICRF3 sources and other known potential phase-reference sources to be linked to ICRF3 that are located close to the JUICE position at around the expected time of arrival at Jupiter. The red triangles mark the spacecraft coordinates in the equatorial system, daily between 2031 Jul 20--24. The grid cells are $1^{\circ} \times 1^{\circ}$.}
    \label{fig:ref-sources}
\end{figure}

The phase coherence of VLBI observations is severely limited by the angular separation of the target and the calibrator, so it is important to choose a calibrator source seen as close as possible to the target. The average sky density of ICRF3 objects between $\pm7^{\circ}$ Ecliptic latitudes, where spacecraft in the solar system generally move, is about 1 radio source in 7 square degrees at X-band (and much less favourable at Ka-band). At most of the time, this is sufficient for choosing calibrator sources for conventional phase referencing observations \citep[e.g.][]{1995ASPC...82..327B} where the required target--calibrator separation is within $\sim 1-2^{\circ}$. (See Fig.~\ref{fig:ref-sources} for an example of potential reference source locations.) However, the sky density of the ICRF3 sources is far from optimal for the most accurate relative VLBI astrometric observations employing the method of in-beam phase referencing \citep[e.g.][]{1999AJ....117.3025F,2014ARA&A..52..339R,2020A&ARv..28....6R}. In this observing setup, the target and calibrator sources are located within the primary beam of the single-dish radio telescopes participating in the VLBI observations. For example, the half-power width of the primary beam is approximately $5^{\prime}$ for a small 25-m diameter dish, while only about $2^{\prime}$ for a larger, 64-m antenna at X-band. The advantage of this configuration compared to the traditional ``nodding'' phase referencing is that all sources can be observed simultaneously, without regular telescope re-pointing. Because of the small angular separation and the simultaneous data acquisition, in-beam phase referencing can in principle reach the thermal noise limit in determining the relative astrometric position of the target \citep{2020A&ARv..28....6R}.

Finding suitable in-beam calibrators is usually not a trivial task. The Radio Fundamental Catalog\footnote{\label{fn:rfc}\url{http://astrogeo.org/rfc/}, accessed on 2023.07.04} \citep[RFC, e.g.][]{2021AJ....161...14P}, the most extensive all-sky list of known compact radio source positions observed with VLBI, currently provides about 4 times of the average sky density in the Ecliptic region compared to ICRF3. While efforts are underway to increase the number of compact VLBI-observed sources around the Ecliptic in general \citep{2017ApJS..230...13S,2023ivs..conf..263G}, the local densification of the celestial reference frame with potentially fainter, currently unknown but VLBI-detectable compact radio sources along the trajectory of JUICE -- both in the interplanetary cruise phase but especially during the science operations in the Jovian system -- is inevitable for the highest-accuracy PRIDE measurements.

Local densifications in selected areas require \textit{(i)} identification of potential radio sources that are suitably compact and bright for VLBI detection and \textit{(ii)} determining their accurate position in ICRF. Radio sky surveys with low (arcsec-scale) resolution, such as FIRST \citep[Faint Images of the Radio Sky at Twenty centimeters,][]{1995ApJ...450..559B} and, more recently, VLASS \citep[Karl G. Jansky Very Large Array Sky Survey,][]{2020PASP..132c5001L} offer a good starting point with their extensive radio source lists, even though their observing frequencies ($1.4$ and $2-4$~GHz, respectively) are below the X- and Ka-band frequencies. A blind survey of more than $21,000$ FIRST sources at $1.4$~GHz \citep{2014AJ....147...14D} led to VLBI detection of about $20\%$ of the sample. Notably, sources with point-like optical counterparts in the Sloan Digital Sky Survey (SDSS) were found to be more likely detected with VLBI \citep{2014AJ....147...14D}. Indeed, earlier studies showed that efficient selection criteria can be defined to filter the initial source list for VLBI-detectable compact objects. Pilot observations of smaller samples of FIRST sources \citep{1999NewAR..43..629G,2008A&A...477..781F} revealed that a significant fraction (up to $85\%$) of the pre-selected sources show mas-scale compact structures at $5$~GHz if total flux density, arcsec-scale compactness, and the existence of an SDSS optical quasar counterpart are taken into account. Nowadays, when VLASS data are available, radio spectral index information can also be incorporated in source pre-selection methods, potentially further improving the efficiency of finding suitable, yet unknown VLBI calibrators. The ultimate proof of the correct pre-selection is the actual VLBI detection of the sources identified as potential calibrators. To this end, prior pilot VLBI observing campaigns targeting the selected new calibrator candidates will be essential to conduct, to cover the celestial areas visited by JUICE in high-priority mission phases in the context of PRIDE operations. Such experiments would also allow us to link these sources to ICRF. This would ensure the most accurate determination of spacecraft position by means of VLBI phase referencing.

\section{PRIDE scheduling}
\label{s:schd}

Scheduling in the VLBI context means providing all radio telescopes participating in an observation with very detailed 'instructions' on where to point and what configuration of telescope instrumentation to engage. It also contains a set of parameters critically important for the data correlation and post-processing. All in all, scheduling is the most important step in conducting a VLBI observation in which the observer exercises the full control over the experiment. For PRIDE proposes, scheduling of major Earth-based experiment assets described in subsection~\ref{ss:PRIDE-network} involves components, typical for any traditional astronomical VLBI and specifics defined by the near-field geometry and other characteristics of spacecraft as a VLBI target. The following subsections describe PRIDE scheduling and its interface with the JUICE mission.

\subsection{PRIDE--JUICE scheduling interface}
\label{ss:sched-m}


The initial step of the PRIDE implementation was to identify science opportunities at various phases of the mission in concurrence with the overall mission science operations plan \citep{lorente2017esa}. 

PRIDE scheduling is based on the planning files provided by the SOC at the different levels of science planning. 

\begin{itemize}
\item At strategic science planning level the intended downlink windows are inferred from a web-based mission timeline \footnote{\url{https://juicesoc.esac.esa.int/tm/?trajectory=CREMA_5_1_150lb_23_1//}, accessed 2023.06.15.}, displaying the trajectory science segmentation defined with the Science Working Groups under the supervision of the Science Working  Team. At this level of planning, the downlink windows schedule remains indicative, and remains to be confirmed. The analysis of the coverage of the strategic planning schedule, from a PRIDE point of view, is made by the team using the SPICE \footnote{\url{https://www.cosmos.esa.int/web/spice//}, accessed on 2023.06.20.}  \citep{costa2018spice} (Spacecraft, Planet, Instrument, "C-matrix," Events tool) libraries, as well as the Planetary Coverage \footnote{\url{https://www.cosmos.esa.int/web/spice/about-webgeocalc}, accessed on 2023.06.20.}  libraries. All files required as input for this coverage analysis (SPICE Kernels) and mission event files \footnote{\url{https://juicesoc.esac.esa.int/event_tool//}, accessed on 2023.06.15.} (moon flybys, station visibility and downlink events, etc.) are produced and provided by the SOC team via the agreed interfaces. 

\item At tactical science planning level (also referred to as medium and short term planning level), a few weeks before the actual execution of spacecraft operations, PRIDE infers the information necessary for its scheduling from a direct access to the final downlink schedule as well as to the different instruments planning files. 
\end{itemize}

\begin{figure}[!htb]
     \centering
     \includegraphics[scale=0.25]{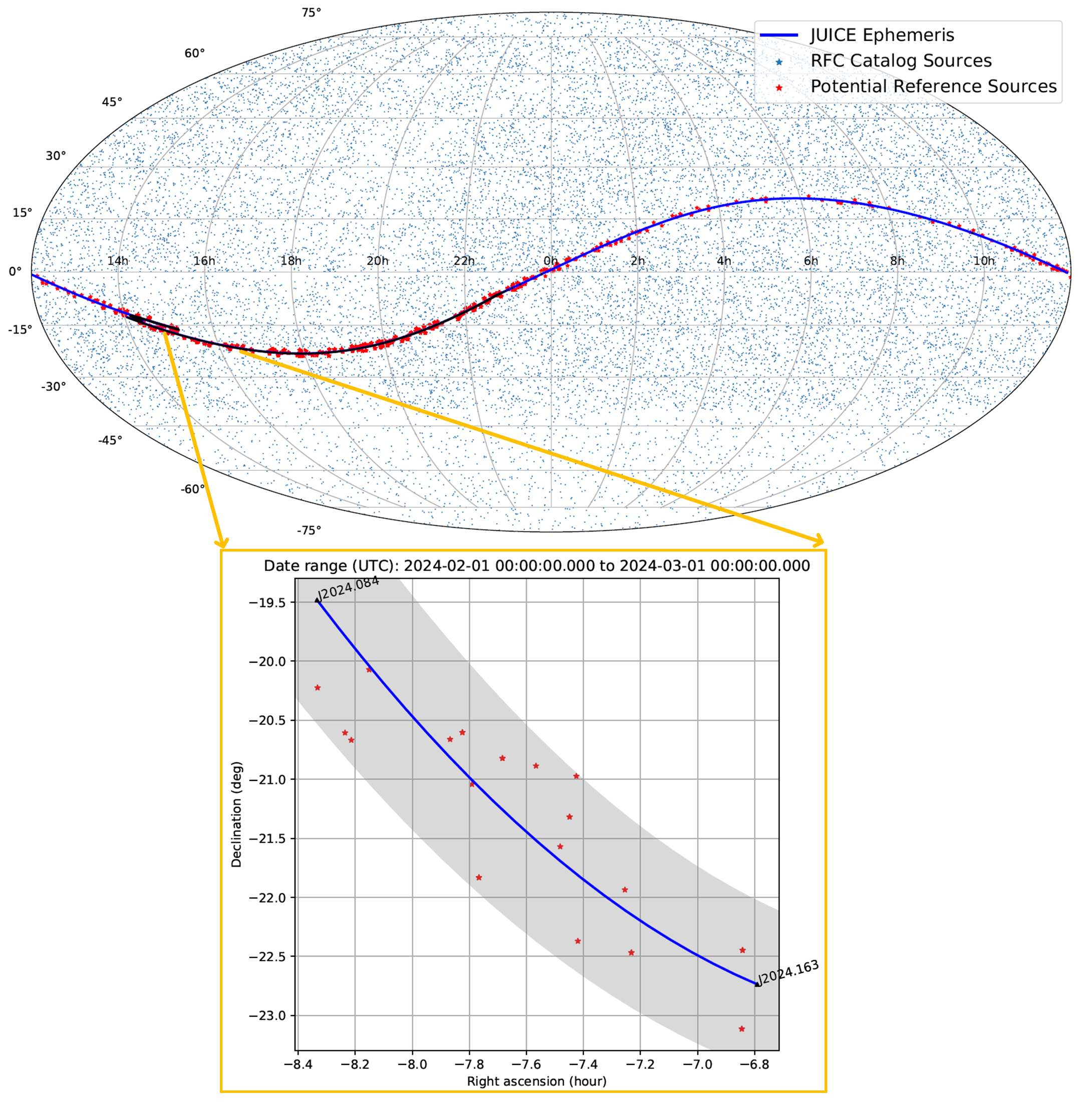}
      \caption{PRIDE observational planning chart for the cruise phase of the JUICE mission (CReMA 5.0) for the time period of 2024.01.01 to 2024.12.31. The upper panel shows the whole sky in the Mollweide projection populated with the RFC celestial sources (blue dots). The lower panel (framed in orange color) shows a zoomed-in finding chart for a period of one month. Red dots indicate potential reference sources from the RFC that are within one degree of the JUICE spacecraft's ephemeris. Blue lines indicate the JUICE celestial track as per CReMA 5.0. The shadowed area indicates the celestial positions within $1^{\circ}$ from the JUICE celestial track. In general, the calibrators are distributed uniformly along the JUICE spacecraft sky track.}
    \label{fig:finding}
 \end{figure}
 
The subsequent step involves organising, planning, and preparing for a PRIDE observation. To observe spacecraft signals with VLBI antennas, an accurate time schedule must be used. This includes downlink and uplink sessions, UT range, and the projected location of a spacecraft at certain epochs. The flux density and compactness of a calibrator (phase reference) source near the target source impact the quality of a VLBI observation, as was described in section ~\ref{s:ph-ref-catalog}. A finding chart (Fig.~\ref{fig:finding}) serves as the starting point for identifying potential phase referencing calibrators near the celestial position of the spacecraft at a given epoch. For PRIDE observations, finding charts are produced using {\tt Spicepy} \citep{altobelli2016science} and {\tt Astropy}  \citep{robitaille2013streicher} software. Finding charts are populated by potential calibrator sources from the Radio Fundamental Catalogue \cite[][see also footnote \ref{fn:rfc}]{2021AJ....161...14P}. The software modules mentioned above and their complete descriptions are provided in \cite{Vidhya+2023ASR}. The next step is to create a one degree wide strip of sky in the center of the ephemeris and search for radio sources within this strip. All of these sources could be used as potential calibrators for the PRIDE experiment, as shown in Fig.~\ref{fig:finding}. 

\subsection{PRIDE scheduling for Earth-based radio telescopes}
\label{ss:sched-vlbi}

Once the most suitable observing window has been defined and the nearby calibrators identified, the next step is to determine the radio telescopes that can observe the spacecraft at the planned time. The participating radio telescopes are selected based on their location and capabilities (\emph{e.g.}, elevation limits, frequency availability, access arrangements) to take into account the spacecraft's visibility and observing time intervals.

Scheduling a VLBI experiment is a process by which each radio telescope's receiving system is configured; setting the desired observing frequency, radio telescope pointing, and start time to record data. The scheduling of PRIDE observations is the same as a typical VLBI session. In standard 'astronomical' VLBI observations, however, the coordinates of a source are considered to be constant during the observing run. In near-field VLBI, the spacecraft can move rapidly across the primary beam of a telescope. Therefore, the position of a spacecraft is calculated at different epochs as a function of time and is then used for repointing the antenna as required by the moving target. There are several programs involved in finalising the schedule. \texttt{MakeKey}\footnote{\url{https://gitlab.com/gofrito/makekey}, accessed 2023.07.04} is a script used to calculate the coordinates of the spacecraft based on the latest JUICE SPICE kernels, provided by SOC. The output of \texttt{MakeKey} is the input for \texttt{pySched}\footnote{pySCHED (\url{https://github.com/jive-vlbi/sched}) is built upon the NRAO SCHED (\url{http://www.aoc.nrao.edu/software/sched/index.html}), both accessed on 2023.07.04.}, a program that produces the standard files that need to be distributed to the different telescopes and correlators for VLBI observations and operations.

\section{PRIDE contribution to state vector estimations}
\label{s:state-estimations}


The precise reconstruction of a spacecraft's trajectory from the tracking data acquired during a space mission allows for the estimation of various parameters influencing the spacecraft dynamics. This includes the determination of the targeted bodies' ephemerides and, for the Galilean satellites and Jupiter, of the parameters characterising the tidal dissipation mechanisms in the Jovian system \cite{Lainey+2009}. The unprecedented improvement in moons' ephemerides expected from the JUICE mission will thus deepen our understanding of planetary system's evolution and will bring new insights into the dynamical history of the Solar System \cite{peale1999,greenberg2010,heller+2015}.

However, the strong dynamical coupling of Io, Europa and Ganymede induced by the Laplace resonances between these moons adds unique challenges to the estimation process \cite{Dirkx+2017,Fayolle+2022PSS}. It would indeed ideally require an evenly distributed data set to obtain a balanced solution and fully benefit from the high accuracy expected for JUICE tracking data \cite{cappuccio+2020,cappuccio+2022,Magnanini+2023}. This strengthens the need for a diversified and synergistic set of observations, to which PRIDE measurements will directly contribute.

\subsection{Spacecraft state vector estimation}
\label{s:state-v}

PRIDE VLBI products will provide complementary information on JUICE's lateral position in the ICRF. It can be treated independently or in combination with other measurements, e.g., provided by the mission nominal orbit determination means \cite{Tanco+2023SSR} or radio science range and range-rate measurements by 3GM \cite{Iess+2023SSR}. The spacecraft's dynamics with respect to the Galilean moons (for the case of flybys and the orbit phase) will mostly be determined from 3GM closed-loop Doppler measurements, possibly supplemented by PRIDE's \textit{ad hoc} open-loop Doppler products. On the other hand, the complementarity between range and VLBI observations plays an important role in the estimation of the moons ephemerides (Section \ref{s:ephemer}), by directly constraining the three-dimensional position of JUICE w.r.t. the Earth. For specific (face on) orbital geometries of JUICE w.r.t. Earth, the PRIDE VLBI data may contribute to the spacecraft's orbit determination, but this is expected to be a rare situation.

\subsection{Natural satellites ephemerides} \label{s:ephemer}

In combination with an accurate planetary ephemeris and the spacecraft's state vector determined from Doppler data, data points on the position of the Jovian satellites can be created from PRIDE data products acquired during the flybys and Ganymede orbit phase. For the case of the JUICE mission, there are a number of aspects that will need to be addressed to make optimal use of these data. The iterative procedure outlined above, where the planetary ephemeris is used as input to the satellite ephemerides will no longer be directly applicable. At present (including several years of Juno data), the incompatibility between DE (JPL Development Ephemeris models, \cite{Park+2021AJ}) and INPOP (Int\'{e}grateur Num\'{e}rique Plan\'{e}taire de l'Observatoire de Paris, \cite{INPOP-2020}),  Jupiter ephemerides is at the level of several kilometers, which is an indication of their current accuracy \cite{fayolle+2023a}. The JUICE tracking data will allow this uncertainty to  be reduced significantly. However, the data that are used to improve the Jupiter ephemeris is also to be used for satellite ephemerides determination. As a result, the data should ideally be used for a concurrent satellite and planetary ephemeris estimation or, at the very least, an iterative procedure of satellite and planetary ephemeris determination. This process is further complicated by the possible need to do a coupled spacecraft-satellite state determination. The underlying reason for the strong need for coupling the estimations of the spacecraft, satellite and Jupiter ephemeris lies in the large mismatch in a priori state uncertainty of the Jovian system and the post-JUICE uncertainty that could be obtained. This fact, in combination with the complications posed by the dynamical coupling of the Galilean moons, makes the decoupling of the different bodies' estimations potentially unfavourable. Different analysis strategies, partially decoupling parts of the solution, will be investigated \citep{Fayolle+2022PSS}. A broader view of this issue is provided by \cite{VanHoolst+2023SSR}. This process may also require the direct linking of previously independent analysis tools to perform the analysis and obtain realistic error bounds on the resulting solution. Combining the JUICE radio data with the existing astrometric data sets used by \cite{Lainey+2009} was analysed by \cite{fayolle+2023b}, where it is shown that the combination of short- and long-term data sets has the potential to stabilise and improve the solutions for the satellite ephemerides and associated dissipation parameters. This further motivates the need to link and integrate existing tools.

The contribution of PRIDE VLBI data to the JUICE-only solution was first quantified in \citep{Dirkx+2017}, in an extended sensitivity analysis parsing various possible tracking and estimation setups. PRIDE VLBI measurements are expected to mostly improve the determination of the Galilean satellites and Jupiter's normal positions (out-of-plane direction), especially for Callisto. Limited VLBI data can indeed be acquired for Io and Europa as no flyby will be performed at the former and only two at the latter. On the other hand, the solution for Ganymede estimated from 3GM range and Doppler measurements is already extremely accurate due to JUICE's orbital phase. Callisto's state solution thus offers the best opportunity for improvement. 

Compared to the analysis performed in \citep{Dirkx+2017}, the current overlapping between the timelines of the JUICE and Europa Clipper missions must now be taken into account, based on their most recently updated Jovian tours \footnote{\url{https://spiftp.esac.esa.int/data/SPICE/JUICE/kernels/spk/}, accessed 2023.07.04}$^{,}$\footnote{\url{https://naif.jpl.nasa.gov/pub/naif/EUROPACLIPPER/kernels/spk/}, accessed 2023.07.04}. Nine JUICE flybys were also added at Callisto since the CReMA 4.2 version used in \citep{Dirkx+2017}. The assessment of PRIDE products' contribution to the ephemerides solution should be revisited accordingly, and rely on the potential improvement that PRIDE data could bring to a joint JUICE--Europa Clipper estimation. In addition, the methodology used in this previous study should be extended by incorporating the spacecraft state vector estimation directly into the estimation, ideally using a coupled estimation \citep{Fayolle+2022PSS}.

The determination of both spacecraft dynamics and natural satellite ephemerides from PRIDE data will be done using a combination of tools. Current ephemerides of the Galilean (and other) satellites determined at the Institute of Celestial Mechanics and Ephemeris Calculations (IMCCE) are created using the NOE software \cite{Lainey+2009}. For ephemerides incorporating the radiometric tracking data in the methodology described by \cite{Fayolle+2022PSS}, the open-source TU Delft Astrodynamics Toolbox (Tudat) software\footnote{Documentation: \url{https://docs.tudat.space}, Source code: \url{https://github.com/tudat-team/}, accessed 2023.07.04} \citep{dirkx2022open}  developed at TU Delft will be used in conjunction with Numerical Orbit and Ephemerides (NOE) software (and possibly other tools). In the analysis performed by \cite{fayolle+2023b}, the numerical orbital propagation of the Galilean satellites of NOE and Tudat was benchmarked, resulting in sub-meter differences over a period of several years. Thus far, Tudat has been used primarily for simulation studies involving a broad variety of tracking and data analysis schemes for the JUICE mission and (Galilean) satellite ephemerides \citep{Dirkx+2016,Dirkx+2017,fayolle2021analytical,Fayolle+2022PSS,fayolle+2023a,fayolle+2023b,villamil2021improvement,plumaris2022cold}. It has been applied to limited analysis of real planetary tracking data \citep{bauer2016demonstration,bauer2017analysis}, and development is currently funded and underway to allow it to read and process both PRIDE Doppler and VLBI data, and typical closed-loop deep-space tracking data. Due to Tudat's open-source nature, all our determinations of spacecraft orbits and associated parameters of interest will be published along with the full analysis code, allowing the entire community to reproduce, scrutinize, and improve upon our work. Through this effort, we will ensure that our data and its associated science products can reach the broadest audience, and allow a much broader community to involve themselves (indepently or in collaboration) in the analysis.

\subsection{PRIDE measurements as a validation data set}

The primary application of the Doppler and VLBI data acquired by PRIDE has been described in Sections \ref{s:state-v} and \ref{s:ephemer}. However, in addition to its direct contribution to the solution of spacecraft, moon and planetary dynamics and associated parameters, the PRIDE data can also play an important role in validating independent solutions and other tracking data. 

As discussed by \cite{Bocanegra+2018}, past experience with the PRIDE open-loop Doppler data has shown that the open-loop Doppler data can, in some cases, be more accurate than the standard tracking, at least for the case of X-band tracking of Mars Express. For JUICE, the situation will be different, since the use of the combined X- and Ka-band links, as well as advanced tropospheric noise characterisation, will significantly improve the closed-loop Doppler tracking data quality acquired by 3GM \cite{cappuccio+2020,cappuccio+2022}. Future experiments in Ka-band, and combined X- and Ka-band tracking, using BepiColombo, will provide insight into the capabilities of the existing PRIDE analysis and processing pipelines to exploit the dual-frequency tracking link. An improvement of the PRIDE Doppler data quality may be obtained through further analysis of the cross-correlation process, since any noise that does not correlate between different stations is a result of non-common noise sources, such as downlink tropospheric, ionospheric and mechanical noise at the receiving station. With such an approach, the PRIDE Doppler data may also reach a level of accuracy where it could be used to validate the data quality and calibrations of the regular closed-loop tracking data.

In addition to providing validation capabilities for the observations, PRIDE data can also be used to validate the ephemeris solutions that are obtained. Conceptually, this is a situation that is similar to the role that satellite laser ranging plays in the orbit determination of \textit{e.g.} the NASA's GRACE (Gravity Recovery and Climate Experiment) spacecraft, where it is used primarily to validate the quality of the high-accuracy spacecraft orbit based on GNSS (Global Navigation Satellite System) measurements. The satellite ephemeris solutions that will be generated during and after the JUICE mission will require data fusion at a level that has not been attempted before \citep{Fayolle+2022PSS}, combining radiometric tracking data from the JUICE and Europa Clipper data, optical astrometry from both missions using the JANUS (Jovis, Amorum ac Natorum Undique Scrutator) camera on JUICE and possibly stellar occultations using the mission's Ultraviolet Spectrograph (UVS) instruments \cite{VanHoolst+2023SSR}, as well as Earth-based astrometry. The attainable quality of the ephemerides that could be achieved is significantly higher than what is currently available. Consequently, deficiencies in the dynamical models that have thus far remained well below the noise floor of the true ephemeris uncertainty will likely become relevant. Unknown errors and issues in both the data fusion and the dynamical modelling will manifest itself as true errors that are (much) bigger than the formal errors. PRIDE can be used as an independent validation of the solution, and a quantification of the true error sources. Specifically, by generating an ephemeris solution \textit{without} the PRIDE data, and subsequently checking the (mis)match between the model and the data, the true to formal error ratio can be quantified. Such a process will be important in improving modelling of the data and the dynamics, which in turn is an essential ingredient of achieving a global, coupled, solution of the ephemerides.

\subsection{Archiving strategy of PRIDE measurements}
\label{ss:archiving}

PRIDE does not get any raw instrument data from the spacecraft thus the definition of PRIDE raw data is different from other experiments. PRIDE raw data are radio signals recorded by Earth-based instrumentation at VLBI radio telescopes. The ground-based array of telescopes participates in a VLBI observation and sends data to the processing centre for specialised processing, the correlation (Section \ref{ss:nfVLBI}). Before the correlation, a typical amount of data from a single PRIDE observing run is of the order of 10--50~Terabytes. In the course of the JUICE mission, PRIDE observations are expected to collect tens of petabytes of telescope raw data. From these VLBI raw data, new datasets are extracted for performing the interferometric and Doppler analysis. For the purpose of the PRIDE experiment, the final dataset after the correlation is considered to be the raw data (in the context of JUICE mission), formatted in FITS (Flexible Image Transport System). The storage needed for the correlated data in FITS format is of the order of hundreds of Gigabytes. For the main deliverable of PRIDE, the correlated data in the form of FITS files are the raw data of the experiment. These datasets allow the production of VLBI maps of the observed patch of the sky containing the spacecraft and are used to extract lateral positional components of its state vector.

Together with the interferometric data, the raw telescope data contain also the observed spectrum of the radio signal transmitted by the spacecraft. As mentioned, these files are not archived because of the large amount of storage needed. Therefore, the outputs of PRIDE high-resolution spectrometer software (Section \ref{ss:PRIDE_Dop}) are stored instead. The spectrometer allows the initial detection of the carrier tone of the spacecraft and the determination of the temporal evolution of its frequency over the entire scan. Binary spectra for each scan of every observing radio telescope are stored in FITS format.

Thus, the raw PRIDE data to archive are the FITS files of the interferometric PRIDE observations and the binary files in FITS format containing JUICE's radio signal spectra.

With regard to the calibrated datasets, the data reduction of the VLBI FITS files is performed with the software CASA (see Section \ref{ss:nfVLBI}), which allows calibrating the VLBI data for atmospheric, geometrical and instrumental effects. The calibrated data consists of the calibrated VLBI datasets in FITS format. From the binary spectra, the spectrometer allows the spacecraft carrier frequency determination at the mHz level and produces ASCII tables containing the time of the observation, the frequency determination and its noise. These files are created and stored for each observing scan of every radio telescope.

Finally, the derived VLBI data are the spacecraft's position in the plane of the sky. This is the main measurable of PRIDE. For this purpose, a table in ASCII format with the time of the observations, right ascension and declination and their errors is stored. Moreover, a table in ASCII format of the topocentric Doppler determination and respective error is also produced. Both tables are later used for ephemerides studies in scientific publications.

For the purpose of data provenance, calibration data will be also provided. The VLBI datasets are written in FITS format. These files contain the tables used for the data reduction in the CASA software. These include flag tables, antenna and phase calibration, and the imaging process parameters. Also, the calibrated VLBI images of phase-referencing calibrators (natural background radio sources) are provided as FITS files. The Doppler analysis does not require calibration. However, the initialisation parameters and input files used in the Doppler processing are provided together with the outputs of the intermediate steps. The input files are stored in ASCII format and the binary output of the intermediate steps in FITS format. Furthermore, all the scripts used for producing the archived tables, plots and images are provided for archiving.

The archiving will follow the Science Data Generation, Validation and Archiving Plan of JUICE. PRIDE data will be available in the ESA PSA in PDS4 standard format after its proprietary period. The EVN proprietary period is one year after the correlated data are distributed to the Principal Investigator. Since the PIs of the VLBI observations are also the PIs of the PRIDE experiments, the proprietary window could be aligned with the ESA standard of 6~months from data release from the correlation center at JIVE.

\section{Multi-spacecraft VLBI observations with PRIDE}
\label{s:Eu-Clip}


In the early 2030s, in addition to the JUICE mission, NASA's Europa Clipper mission will perform science operations in the Jovian system, with a focus on Europa science. As discussed in Section \ref{s:ephemer}, the Galilean satellite ephemerides will be much improved by combining the Clipper radio science \citep{mazarico2023europa} with the JUICE data, as well as with historical data sets. A comprehensive simulated analysis of the combined JUICE-Clipper radio science data set for the determination of geodetic parameters of interest and ephemerides is provided by \cite{Magnanini+2023}. The analysis by \cite{fayolle+2023b} also uses both missions' radio tracking data sets in their analysis, to ascertain the influence of combining astrometric and radiometric data sets for Galilean satellites. These analyses both indicate that the combination of data sets will greatly strengthen the quality of the resulting science data products. For PRIDE specifically, however, there is an additional unique benefit to having JUICE and Clipper in the Jovian system at the same time: the opportunity to perform concurrent observations.

Specifically, the synchronicity of the JUICE and Clipper missions offers the unique possibility to track the signals of both spacecraft within the primary beam of the telescope, generating so-called multi-spacecraft in-beam VLBI measurements \citep{2010evn..confE..66F,Molera+2021}. In beam-tracking does not require nodding between the target and a reference phase source (or between two targets in case of multi-spacecraft observations). The reduced spatial and temporal differences in the paths of the two signals tracked simultaneously thus cause most systematic errors to cancel out \citep{MajidBagri2008}, such that highly accurate VLBI measurements can be expected. Previous tracking experiments between the Phoenix probe and two Martian orbiters (Odyssey and Mars Reconnaissance Orbiter) indeed showed that the accuracy of VLBI observations can be expected to get lower than 0.1\,nrad in case of in-beam tracking \citep{2010evn..confE..66F}. The capability to perform such multi-spacecraft in-beam observations was moreover specifically demonstrated for PRIDE in 2019 by simultaneously tracking several Martian orbiters and landers (Mars Express, InSight, Odyssey, MRO, ExoMars Trace Gas Orbiter) \citep{Molera+2021}.

To realise in-beam observations, the two spacecraft should both be transmitting and their celestial angular separation should be small enough to be within the primary beam of the telescope (without nodding between the two targets). In addition to the mere technical feasibility, other considerations should be discussed. For the JUICE and Clipper missions in particular, choosing to perform in-beam observations of the two spacecraft shortly after and/or before they each perform a flyby (or orbit) at a different moon could yield critical constraints on the relative positions of the Galilean satellites and thus help solving their highly coupled dynamics (see Section \ref{s:state-estimations}). The relative angular position that can be obtained in this manner can also be achieved using optical data during so-called mutual events. However, the relative position using VLBI observations would allow such observables to be obtained at an accuracy that is order of magnitude more accurate (provided the spacecraft trajectories are compatible) and the opportunity provided by the concurrent in-system missions should be exploited to the full extent.

\begin{figure}[h!]
    \centering
    \includegraphics[width = 1.05\columnwidth]{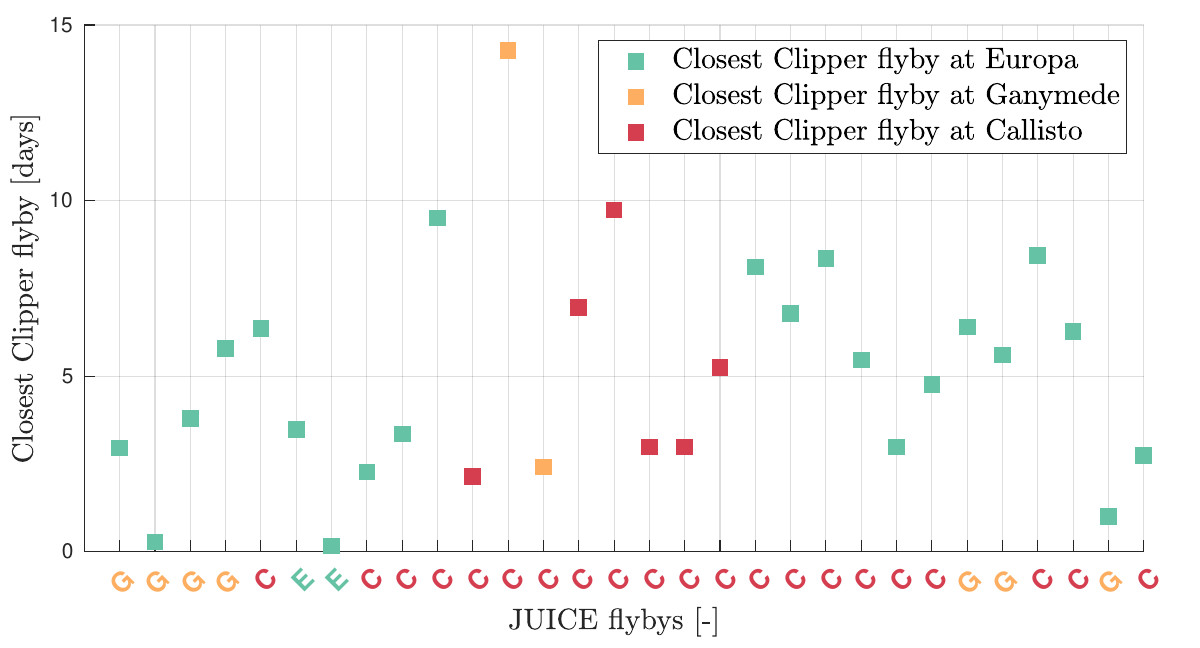} 
    \caption{Closest combinations of JUICE's and Clipper's flybys. The flyby moons are colour-coded and the time gap between each JUICE flyby and the nearest Clipper flyby (temporally speaking) is displayed on the y-axis.}
    \label{fig:closestFlybys}
\end{figure}

Based on the latest versions of JUICE and Clipper trajectories, several flyby combinations appear to fulfil the previous requirements, as shown in Fig. \ref{fig:closestFlybys}. In particular, we can identify 7 combinations of JUICE and Clipper flybys performed at different moons within three days of  each other. These configurations will be further investigated in a dedicated study.

The contribution of such in-beam measurements to the global estimation solution might however be limited if all range, range-rate, and PRIDE VLBI data are combined and processed together, especially if both the JUICE and Europa Clipper missions are considered. The expected precision level for multi-spacecraft in-beam measurements (0.1 nrad in relative angular position \cite{2010evn..confE..66F}) indeed corresponds to ~60-100 m at Jupiter's distance, which is larger than most uncertainties in the moons' positions predicted by a joint JUICE-Clipper solution \cite{Magnanini+2023}.

Nonetheless, such in-beam multi-spacecraft observations represent invaluable validation opportunities for the solutions independently determined from JUICE and/or Clipper other tracking data. In particular, we can verify that these local measurements of the two spacecraft's relative angular position are consistent with the boundaries defined by the global solution's formal uncertainties. Any discrepancy will be extremely helpful to detect, identify and possibly mitigate potential dynamical modelling issues. For the JUICE mission, dynamical modelling for the spacecraft and/or the natural satellites represent a major challenge if we hope to reach the very low uncertainty levels predicted by covariance analyses for the moons' ephemerides and associated dynamical parameters \cite{Fayolle+2022PSS,Magnanini+2023}. Validation of the estimated solution(s) will thus be critical and can be greatly supported by simultaneous measurements of both spacecraft independent of range, Doppler and classical single-spacecraft VLBI data.

\section{Radio occultation observations with PRIDE}
\label{s:occult}

\begin{figure}[h!]
    \centering
    \includegraphics[scale=0.5]{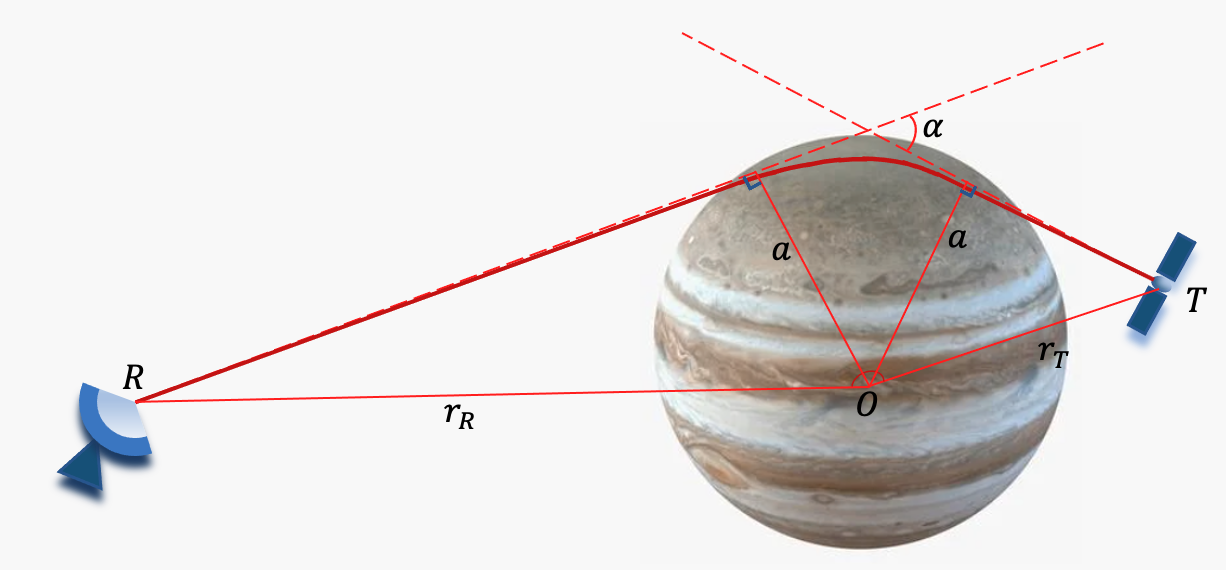}
    \caption{Schematic representation of the geometry of the radio occultation experiment. The signal transmitted by the spacecraft $T$ gets occulted by the planet's atmosphere in its trajectory to the receiver $R$ on Earth. As the signal gets refracted by the atmosphere (or ionosphere) it is bent by an angle $\alpha$ at an impact parameter $a$. By determining these two parameters, also known as the ray path parameters, information such as density, temperature and pressure of the probed vertical atmospheric profile can be determined.}
    \label{fig:rad_occ_geom}
\end{figure}

PRIDE radio occultation measurements will be conducted as JUICE passes behind the limb of the occulting body. Below we briefly discuss such the PRIDE occultation measurements of Jupiter,  Ganymede and the rings of Jupiter. In a radio occultation experiment, the transmitting signal by the orbiting spacecraft experiences refraction and absorption as it passes behind the visible limb of the planetary body due to its propagation through the planet’s atmosphere and ionosphere, and scattering as it traverses through the planetary rings, on its way to the Earth-based tracking stations (Figure \ref{fig:rad_occ_geom}). By analysing the changes in phase and amplitude of the received carrier signal, physical properties of the planetary atmosphere and ionosphere can be inferred, such as density, temperature, pressure and abundance of chemical compounds in the atmosphere, and the size and distribution of particles in the rings.

The PRIDE-JUICE radio occultation experiment will complement the radio occultation experiment conducted by 3GM (with Estrack) by receiving the spacecraft signal with multiple radio telescopes around the world in one-way (or three-way) mode at X- and Ka- bands. During a radio occultation experiment, the signal-to-noise ratio (SNR) of the received signal is of particular importance, since for instance, it will determine the penetration depth of the signal in the atmosphere. In the PRIDE-JUICE setup, the use of a large number of antennas, and of large aperture antennas in particular (e.g., 64-m Tianma (China), 64-m Sardinia (Italy), 100-m Effelsberg (Germany), 110-m Green Bank (USA) and 500-m FAST (Five-hundred-meter Aperture Spherical radio Telescope, China), can improve the SNR \cite{Bocanegra+2018} and hence the quality of the received radio occultation signal, allowing for a more precise derivation of atmospheric data and sounding deeper in the atmosphere. High SNR radio occultation signals will, for instance, allow a better characterisation of NH$_3$ and PH$_3$ abundances, as well as condensation in hazes and clouds, as the signal probes deeper in the atmosphere and could enable the characterisation of the tenuous Jupiter rings, previously undetected by other Jupiter radio occultation experiments \cite{Tyler+1981}.


\subsection{Jupiter’s hazes and clouds}
\label{ss:hazes-clouds}

The distribution of gas phase species (H$_2$O, H$_2$S, NH$_3$) as well as their condensates are driven by the dynamics of clouds formation and transportation of gases that photo-dissociate with solar UV photons and produce tropospheric and stratospheric hazes. Jupiter exhibits axisymmetric bands that can be warm, cyclonic `belts' and cool, anticyclonic `zones’ presenting different temperature gradients and zonal jets \cite{Flasar+2004,Fletcher+2016Icar,Fletcher+2023SSR}. Aerosol properties can be different in these different bands. In the bright zones, volatile species such as ammonia become saturated and condense to NH$_3$ ice at the colder temperatures of the zones, while in other zones aerosols sublimates in the warmer and typically cloud-free belts. Following the properties, distribution and variability of hazes and clouds of the atmosphere of Jupiter will allow to gain insights on the vertical and horizontal dynamics of the troposphere. The retrievals of NH$_3$ and PH$_3$ abundances by radio occultation will lead to better constraints for the characterisation of  processes such as condensation, photo-dissociation as well as chemical reactions occurring in Jupiter’s atmosphere, and provide better constraints on the photo-chemical processes occurring in the different zones. These observations will probe down to approximately the cloud level at the 700\,mbar level, at which NH$_3$ absorption fully attenuates the signal. The X-band radio occultation measurements will allow probing deeper in the atmosphere than Ka-band ($\sim$500\,mbar), as the latter is affected by a larger refractive defocusing loss and critical refraction limit. The penetration depths were estimated simulating the expected total signal attenuation received by the radio telescopes on Earth at the distance of Jupiter during a radio occultation experiment, at X- and Ka-band, and comparing it with the expected absorptivity profiles determined from nominal Jupiter atmospheric composition models.



\subsection{Jupiter’s stratosphere}
\label{ss:J-strat}
The stratospheric circulation of Jupiter also shows zonal organisation as observed for the troposphere, with bands of warmer and cooler regions \cite{Antunan+2021NatAs}. This zonal organisation is observed to be important at low latitudes.  At the equator, the stratosphere shows oscillations of the 10\,mbar thermal contrasts that vary on a 4-year timescale \cite{Leovy+1991Nat,Orton+1991Sci}. The vertical structure of this pattern, is thought to be driven by waves from the troposphere that are interacting with  the mean flow \cite{Friedson-1999Icar}. With radio occultations, the vertical structure of the atmosphere and its interaction with the deeper layers will be observed and will be followed over the duration of the mission, showing the evolution on 4-years period. 

\subsection{Ganymede's ionosphere} 
\label{ss:Gan-exo}
On Ganymede, a clear boundary on the surface can be observed if terrains are located in closed magnetic field lines regions (both ends of field lines from Ganymede intersect with Ganymede's surface) or open field lines regions (one end of the field line intersects with Ganymede's surface and the other end reaches Jupiter). The open field lines region is populated with energetic ions and electrons from Jupiter's magnetosphere and thus the surface is much more exposed to plasma irradiation than on the closed field line regions. Ganymede ionosphere has been recently observed by the Juno spacecraft \cite{buccino2022}. The density of the ionosphere has been determined in ingress, which was in open field lines, but not in egress, in which these lines were closed. These observations in occultation support the fact  that the ionosphere is generated by impact of the surface with ions such as O$^{2+}$ \cite{CARNIELLI2019}. With the PRIDE-JUICE radio occultation, we will be able to follow Ganymede's ionosphere, and more specifically perform direct measurement of the electron content along the propagation path as well and its variation as Ganymede evolves in Jupiter's plasma sheet \cite{Paranicas2018}. Depending on the orbital position of Ganymede on the timescale of the JUICE mission, the flux of ions impacting different terrains on Ganymede and their variations will be reflected by the ionosphere density.

\label{ss:J-sci}


\subsection{Rings}

The Jovian rings have been observed by Voyager 1 and 2, Pioneer 11 and Galileo, as well as by the Hubble Space Telescope (HST) and ground-based telescopes between the wavelengths 0.4--2.2\,$\mu$m \cite{Burns+2004, DePater+2018}. The rings are believed to be composed primarily of dust particles resulting from impacts with ring-moons and collisions between cm- and hundred of meters sized bodies. Nonetheless, these larger-sized particles ($>10\mu$m) have not been individually observed \cite{DePater+2018}. Tyler et al. \cite{Tyler+1981} first implemented the radio occultation technique to detect planetary rings during Voyager 1 flyby of Jupiter. However, this single S/ X-band radio occultation observation was not able to detect Jupiter’s tenuous rings. Given that JUICE will orbit Jupiter for 4 years and has the ability of X- and Ka- band onboard the spacecraft (\emph{e.g.}, sensitive to particles sizes larger than millimeters because of Ka-band availability), we will conduct observations of the spacecraft signal as it gets occulted by the rings and cross-correlate the observed features in the broadband spectrum of the open-loop signal received at multiple radio telescopes on Earth, with the purpose of performing a high-resolution reconstruction of the radial profile of the optical depth to look for the missing `
``parent bodies'' of the dust particles in the rings. PRIDE will be able to reconstruct the opacity profile of the rings occultation comparing the multiple observations obtained a each telescope, enabling a more robust and precise determination of the dips in the apparent radio opacity due to the presence of rings. 


\section{Interplanetary medium diagnostics}
\label{s:plasma}

Interplanetary plasma turbulence is one of the major contributors (along with the Earth's ionosphere and troposphere) to the phase fluctuations of the radio signals between ground stations and spacecraft, thus affecting the accuracy of spacecraft state vectors determination. Studies of the statistical properties of the signal phase fluctuations in their relation with other factors, such as the solar elongation, distance, and solar activity contributes to better interpretation of the results of spacecraft Doppler and VLBI measurements. Such a study in a framework of PRIDE-JUICE will benefit from the standard radio communication system of the planetary mission to investigate the interplanetary plasma. The Interplanetary Medium Diagnostics (IMD) are normally conducted in a 3-way radio link configuration in which the spacecraft is phase-locked to the transmitting ground station via the up-link X- or Ka-band signal while the participating VLBI radio telescopes receive the coherent down-link signal at X- and/or Ka-bands. For the purpose of IMD study, each station can operate in a single-dish mode, meaning the bulk of the broad band data is processed with the Spacecraft Doppler tracking software (see subsection~\ref{ss:PRIDE_Dop}) at each station separately while the reduced narrow band data such as residual (with respect to a-priory model) frequency and phase data are analysed in a combined way. 


The PRIDE team have tracked the radio signals of various planetary mission spacecraft with ground-based radio telescopes to study the relation between the Total Electron Content (TEC) of the solar wind along the line of sight and the Interplanetary Scintillation (IPS) for the past 15 years. Observations of ESA's Venus Express with a number of EVN (and associated) stations were conducted between 2009 and 2014 ~\cite{Molera+2014}. Over 200 observing sessions were arranged utilising 20 radio telescopes in Australia, Asia and Europe. The sessions covered multiple complete Venus orbits around the Sun. The power density spectra of the phase fluctuations at different solar elongations were estimated and show the average spectral index of $-2.41 \pm 0.25$, and essentially independent of the solar elongation while the RMS of fluctuations is basically proportional to TEC which is in a good agreement with the Kolmogorov theory of turbulent media statistical properties.

Further PRIDE observations to improve the IMD-study results continued with the ESA's Mars Express, Rosetta, and more recently BepiColombo planetary missions. We have conducted over 500 sessions for MEX with nearly 40 VLBI radio telescopes. These experiments covered the period between 2014 through 2022, including three full orbits of Mars around the Sun. The data were used for better characterisation of the solar wind and  the space weather forecast~\cite{Molera+2017}. 

\begin{figure}[!hbt]
    \centering
     \includegraphics[width=\columnwidth]{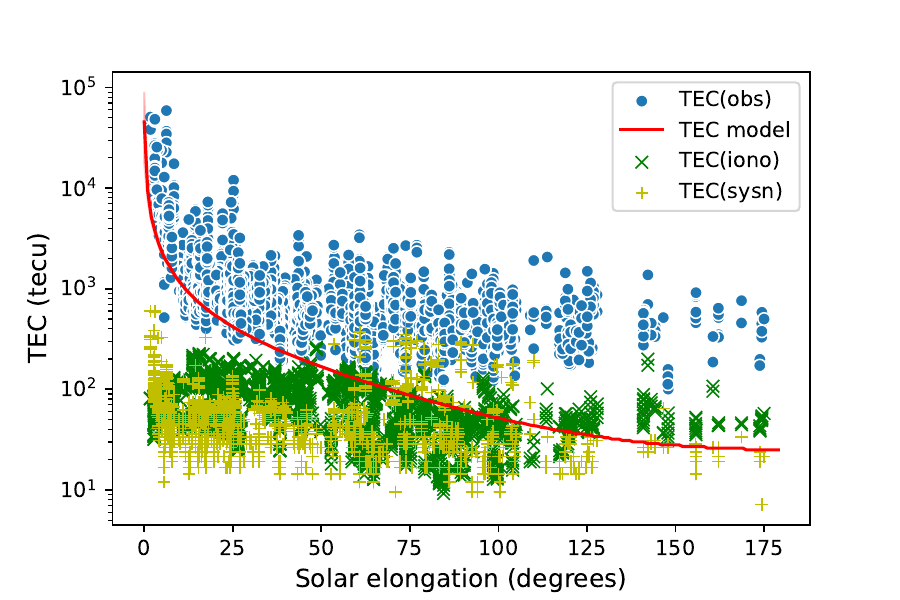}
    \caption{Total Electron Content contributions from ionospheric (both terrestrial and Martian), interplanetary plasma (including solar wind) and system noise, compared to the theoretical TEC model using data from Mars Express mission from 2013 through 2020~\cite{kummamuru+2023}. The observations covered two full orbits of Mars around the Sun.}
    \label{fig:tec}
\end{figure}

The method of characterisation of the solar wind turbulence and interplanetary scintillation for different solar elongations and radial distances were described in~\cite{kummamuru+2023}. We used the detected phase residuals to estimate the total electron content of the solar plasma along the line-of-sight while filtering out the mechanical and ionospheric noise effects. The results of this analysis are presented in Fig.~\ref{fig:tec}. The scintillation spectra indices as determined for all the MEX' sessions are on average at the level of $-2.43\pm0.11$, consistent with the previous studies \cite{kummamuru+2023}. A scaling factor ($\kappa$) was used to obtain the TEC from the phase scintillation measurements ($\rm{TEC}=\kappa\cdot\sigma$)~\cite{Molera+2014}. The best value for $\kappa$ for MEX is $\kappa \simeq 2390$. Figure~\ref{fig:tec} depicts the TEC corresponding to various factors associated with the observations.

The TEC for the line of sight has been empirically modelled with over ten years of collected data. These models are compatible to any future space mission and in particular to JUICE at X-band. The model for the TEC at Ka-band is an ongoing framework of the JUICE mission. Worth mentioning is that dual band (simultaneous X- and Ka-band) observations will provide new material for the study and allow to separate the influence of different dispersive and non-dispersive factors. The latter might provide additional input into plasma studies by JUICE described in \cite{Masters+2023SSR}. These measurements will also allow us to get the contribution into TEC values of local ionospheres of those moons, e.g. Europa and Callisto, to which JUICE spacecraft will be in the occultation or near-occultation (i.e., not occulted by close to the moon's limb as seen from Earth) trajectory configurations. 



\section{Concluding remarks}
\label{s:conclu}

PRIDE is a component of the science suite of the JUICE mission aiming at enhancing the mission's science output by exploiting the spacecraft onboard hardware (essentially -- the radio system) and a global network of VLBI-equipped radio telescopes and data processing facilities. None of these assets is built specially for PRIDE. Due to this exceptionally high reliance on the infrastructure built and operated for other purposes, PRIDE achieves its main goals by providing measurements of the lateral celestial position of the spacecraft and its radial velocity in an ad hoc fashion with respect to the nominal mission operations and within standard practices Earth-based VLBI networks. Such an approach has been successfully demonstrated over the past two decades for several ESA (Huygens, Venus Express, Mars Express) and other planetary and space science missions. As demonstrated in sections~\ref{s:state-estimations}, \ref{s:occult} and \ref{s:plasma}, the PRIDE observing and data handling procedures have been proved in various space missions. 

PRIDE is deeply rooted in the developments of the VLBI technique over the past half a century. The fact that the first ESA Jovian mission JUICE has Jupiter as one of its targets while having in its science suite a VLBI-based experiment, PRIDE, might be seen as highly symbolic: Jupiter was the target of one of the very first VLBI observations in 1967 \cite{Brown+1968ApL}. More than six decades later, the VLBI technique comes back to Jupiter in a new, near-field spacecraft tracking incarnation. 

\bmhead{Acknowledgements} 
The authors express their gratitude to the reviewers and editors for very useful comments, corrections and suggestions.
The observing examples shown in this paper for PRIDE-style tracking of various spacecraft are based on observations conducted with the European VLBI Network (EVN), the AuScope facility, and the Long Baseline Array (LBA), and Very Long Baseline Array (VLBA). The EVN is a joint facility of independent European, African, Asian, and North American radio astronomy institutes. The AuScope involvement in the presented here observations was enabled by Geoscience Australia and the Australian Government via the National Collaborative Research Infrastructure Strategy (NCRIS). The LBA is operated as a National Facility, and managed by CSIRO and the University of Tasmania. The VLBA is an instrument of the National Radio Astronomy Observatory, a facility of the National Science Foundation operated under cooperative agreement by Associated Universities. We are grateful to the ESA personnel for supporting PRIDE observations of the Mars Express and BepiColombo missions. We gratefully acknowledge the VLBI database of the Astrogeo Center maintained by Leonid Petrov (NASA GSFC) and his comments on the contents of the current paper. We thank Chris Phillips (CASS) for supporting PRIDE test observations at the Mopra radio telescope. J.F., S.F., and K.P. acknowledge the ESA PRODEX support (project PEA 4000136207). M.S.F acknowledges partial funding by ESA’s OSIP (Open Space Innovation Platform) program.

\bmhead{Conflict of interest}
The authors declare no conflicting interests, individual or institutional, related to the current paper as a whole and/or any its component. 

\bibliography{biblioPRIDE}  


\end{document}